# Explicit solutions for multi-layered wide plates and beams with perfect and imperfect bonding and delaminations under thermo-mechanical loading.

Marco Pelassa and Roberta Massabò*
Polytechnic School, University of Genova
Department of Civil, Chemical and Environmental Engineering
Via Montallegro 1, 16145 Genova, Italy.

**ABSTRACT**

Explicit expressions, for efficient application in engineering practice, are derived for generalized displacements and stresses in simply supported multi-layered wide plates and beams subjected to steady-state thermal and mechanical loading. The expressions are general and apply to plates composed of an arbitrary number of layers, of arbitrary thickness and elastic/thermal properties, and where the interfaces between the layers may be imperfect and allow relative sliding. The closed-form solutions are obtained using a multiscale homogenized model which depends on only three displacement variables and overcomes limitations of current approaches based on computationally expensive discrete-layer models. The accuracy of the expressions in predicting the highly complex and discontinuous fields, which characterize the response of thick and highly anisotropic plates with interlayer damage and delaminations, is verified using exact 2D thermo-elasticity solutions. The asymptotic limits of the model/solution correspond to the problems of an intact and a multiply delaminated plate. They are derived using a perturbation technique, which also explains the multiscale dependence of the model on the parameters.

## 1. INTRODUCTION

The use of multi-layered composite systems for structural applications has become more widespread over the last decade in many fields of engineering. Multi-layered systems for civil engineering applications include steel-concrete assemblages, laminated wood and cross-laminated timber, where the layers are typically glued or connected by more or less flexible mechanical devices. In aeronautical, aerospace, naval and defense applications, laminated composite systems made of unidirectionally reinforced laminae and 3D woven or through-thickness reinforced laminates are typically used, along with sandwich systems where thin and stiff laminated sheets are connected by more flexible and thicker cores. These systems may be subjected to extreme loading conditions in aggressive environments. Composite laminates and sandwiches, for instance, are currently being pursued for defense applications where the material systems will have to withstand and survive severe loads in extreme environments characterized by very high and very low temperatures [1].

* Corresponding author. Department of Civil, Chemical and Environmental Engineering, University of Genova, Via Montallegro 1, 16145 Genova, Italy. Phone: +39 010 353 2956; Fax: +39 010 353 3546. Email address: roberta.massabo@unige.it

Prediction of displacements and stresses in multilayered systems subjected to thermo-mechanical loading, even under steady-state quasi-static conditions, is complicated by a number of factors. The different mechanical and thermal properties of the layers lead to complex zig-zag distributions of stresses and in-plane displacements in the thickness direction, even in cases where homogeneous systems would deform freely in the absence of stresses, e.g. thermal loading of statically determinate systems [2,3]. The presence of defects, damaged areas or delaminations, which may be due to manufacturing errors, degradation of the adhesives or of the connecting mechanical devices or previous impacts, severely affects the response and modifies the fields; in particular, relative displacements arise at the damaged layer interfaces [4]. In addition, when the conditions become critical the delaminations may propagate leading to substantial reduction of stiffness and often to final failure [5, 6].

In this paper, the term *interface* is used to describe the surface between adjacent layers in laminates or the thin adhesive layers used in multi-layered systems; and *perfect interface* indicates full bonding of the adjacent layers while *imperfect interface* indicates an interface which allows free or controlled relative sliding of the adjacent layers; a *fully debonded interface* describes a delamination.

Accurate solutions for multilayered structures typically require the use of complex discrete-layer models, e.g. layer-wise theories and discrete-layer cohesive-crack models, where the number of unknowns depend on the number of layers and on the kinematic fields assumed in each layer. In the simplest case of a laminated beam described as an assembly of $n$ Timoshenko layers, for instance, the displacement unknowns would be $3 \times n$ [6,7]. This complexity has two drawbacks: first the number of problems which can be solved analytically is limited, as are the closed-form solutions and explicit formulas for stresses and displacements, which are so useful in the engineering practice; and the numerical solutions of the problems are computationally very expensive, especially when the systems exit the elastic regime and the response is dominated by damage progression.

The zig-zag theories originally formulated for fully bonded systems successfully overcome the limitations of the discrete approaches through the definition of homogenized displacement fields depending on a limited number of unknowns, which is equivalent to that of single layer theories [8,9,10,11]. On the other hand, the extension of the zig-zag theories to structures with imperfect interfaces and delaminations has not been successful and early models in the literature give acceptable results only in structures with slightly imperfect interfaces ([12-16]; see also [17] for a discussion). Only recently [17,18] the authors were able to demonstrate the efficacy of the zig-zag approach also for structures with imperfect interfaces and delaminations. They proved that the inconsistencies and limitations of previous models, based on the original ideas in [8,9], to describe systems with imperfect interfaces, were due to the omission of the energy contribution of the imperfect interfaces in the weak form derivation of the equilibrium equations of the problem.

The energetically consistent formulations in [17,18] are limited to plates and beams subjected to mechanical loading and the important thermo-mechanical problem has not yet been solved consistently. In addition, while many applications are presented in [17,18] to highlight the accuracy of the approach for plates composed of different materials, layups and interfacial damage conditions, closed-form expressions for displacements and stresses are not given and the existence of asymptotic limits of the model/solution is observed but not demonstrated.

In this paper, the energetically consistent model formulated in [18] for plates deforming in cylindrical bending and subjected to static and dynamic mechanical loading, is extended to plates subjected to thermal loading (preliminary results in [19]) . The formulation is based on a multiscale approach (see for instance [20,21] for 2D problems) which couples a coarse-grained model (first order shear deformation theory), which describes the global fields, and a detailed small-scale model (discrete-layer cohesive-crack model), which describes the local fields. A homogenization technique is used to average out the small scale variables and define equilibrium equations and fields depending on the global variables only.

The homogenized equilibrium equations are decoupled and solved to obtain accurate and efficient explicit formulas for generalized displacements and stresses in simply-supported wide plates and beams subjected to quasi-static transverse loads and steady state thermal gradients with uniform and sinusoidal distributions. The expressions are general and apply to plates composed of an arbitrary number of layers, of arbitrary thickness and material properties, and where the interfaces between the layers may be fully bonded, fully debonded or partially debonded, and characterized by piece-wise linear interfacial traction laws.

Perturbation theory is used to highlight the multiscale dependence of the model on the parameters and derive the asymptotic limits, which correspond to those of a fully bonded Timoshenko beam and of a stack of fully debonded Euler-Bernoulli beams free to slide along each other. In the fully debonded limit, the zero-order perturbation equations and solution define the global fields, while the higher-order equations and solutions define the small-scale enrichments needed to fully predict the highly discontinuous fields. The perturbation technique also explains the fictitious boundary layers which may arise at the ends of the plate for certain loading and boundary conditions, e.g. clamped edges [22], and are a consequence of a singularity in the model which is introduced by the homogenization technique.

An exact 2D thermo-elasticity solution is derived in the Appendix D for multilayered plates with mechanically and thermally imperfect interfaces and used to verify the multiscale approach. The derivation extends the procedures proposed for fully bonded plates in [2,3,23,24] and is based on the solution of the heat conduction equation and Navier's thermo-elasticity equations.

## 2. MODEL: ASSUMPTIONS AND MULTISCALE APPROACH

### 2.1 Assumptions and definitions

The proposed model refers to the rectangular multilayered plate of Fig. (1), with global thickness $h$ and in-plane dimensions $L_1$ and $L_2 = L$, with $L_1 >> L_2$. The plane $x_3 = 0$ defines the reference surface of the plate in a system of Cartesian coordinates, $x_1 - x_2 - x_3$, whose origin is arbitrarily placed. The plate is subjected to steady-state thermal loads and to mechanical loads acting on the upper and lower surfaces, $\mathscr{S}^+$ and $\mathscr{S}^-$, and on the lateral bounding surface, $\mathscr{B}$, applied so as to satisfy plane strain conditions parallel to the plane $x_2 - x_3$.

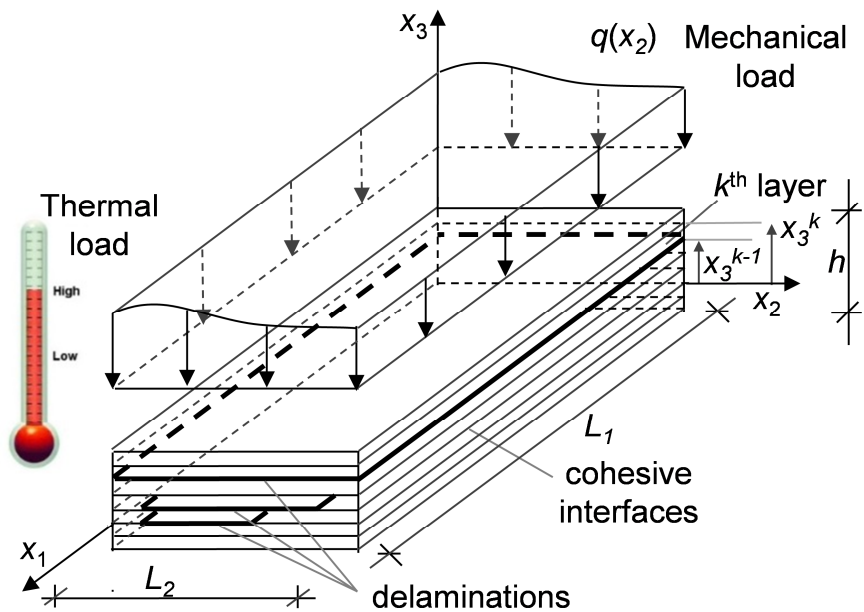

Figure 1. (a) Composite wide plate subjected to thermo-mechanical load, showing discretization into layers, imperfect/cohesive interfaces and delaminations.

The plate consists of $n$ layers exhibiting different mechanical properties and joined by $n-1$ zero-thickness interfaces. The layer $k$, where the index $k=1,...,n$ is numbered from bottom to top, is defined by the coordinates $x_3^{k-1}$ and $x_3^k$ of its lower and upper interfaces, $^{(k)}\mathscr{S}^-$ and $^{(k)}\mathscr{S}^+$, and has thickness $^{(k)}h$. The $k$ superscript in round brackets on the left of a quantity identifies affiliation with layer $k$, while the $k$ superscript on the right identifies the interface. Each layer is linearly elastic, homogeneous and orthotropic with principal material axes parallel to the geometrical axes (e.g. isotropic materials or $0°/90°$ fiber orientation).

The constitutive equations in the layer $k$ are derived from the 3D thermo-elasticity equations [26], which neglect coupling between the heat conduction problem and the elasticity problem and assume prescribed temperature fields. The equations are particularized to a material with no extension-shear coupling ($x_2, x_2, x_3$ are principal material directions) and to plane-strain conditions parallel to the plane $x_2 - x_3$. In addition, the classical assumption of lower-order plate theories is imposed and the transverse normal stresses $^{(k)}\sigma_{33}$ are assumed to be negligibly small compared to the other components and set equal to zero (a more general treatment for 2D plates subjected to mechanical loading, which accounts for the transverse normal stresses is presented in [18]). The resulting equations are:

$$^{(k)}\sigma_{22} = {}^{(k)}\bar{C}_{22}\left({}^{(k)}\varepsilon_{22} - {}^{(k)}\tilde{\varepsilon}_{22}^t\right) \tag{1}$$

$$^{(k)}\sigma_{23} = {}^{(k)}C_{55} 2 {}^{(k)}\varepsilon_{23}$$

with $^{(k)}\bar{C}_{ij} = {}^{(k)}\left(C_{ij} - C_{i3}C_{3j}/C_{33}\right)$, where the $^{(k)}C_{ij}$ are the coefficients of the 6×6 stiffness matrix in engineering notation; $^{(k)}\tilde{\varepsilon}_{22}^t = {}^{(k)}\left(\tilde{\alpha}_2 T(x_2, x_3)\right)$ is the thermal strain component modified to account for plane strain conditions, with $^{(k)}\tilde{\alpha}_2 = {}^{(k)}\alpha_2 + {}^{(k)}(\alpha_1 \bar{C}_{12}/\bar{C}_{22})$ and $^{(k)}\alpha_\alpha$ the 3D coefficient of thermal expansion along the $x_\alpha$ principal material direction; $^{(k)}T = {}^{(k)}T(x_2, x_3)$ is the temperature increment in the layer $k$.

The displacement vector of an arbitrary point of the plate in the layer $k$ at the coordinate $\mathbf{x} = \{x_1, x_2, x_3\}^T$ is defined by $^{(k)}\mathbf{v} = \{{}^{(k)}v_1 = 0, {}^{(k)}v_2, {}^{(k)}v_3\}^T$.

The interfaces, $^{(k)}\mathscr{S}^\pm$ with $k=1,...,n-1$, are mathematical surfaces which are used to describe thin elastic interlayers, damaged regions and delaminations. At the interfaces, material properties and displacements may change discontinuously. The relative sliding displacement (mode II) of the layers $k$ and $k+1$ at the interface $^{(k)}\mathscr{S}^+$ is defined as:

$$\hat{v}_2^k(x_2) = {}^{(k+1)}v_2(x_2, x_3 = x_3^k) - {}^{(k)}v_2(x_2, x_3 = x_3^k), \qquad (2)$$

while the relative opening displacement is assumed to be zero and the interfaces to be rigid against mode I displacements, $\hat{v}_3^k = 0$ [1]. The interfacial tractions, $\hat{\sigma}_S^k = [{}^{(k)}\sigma_{23}(x_3 = x_3^k)]n_3^k\big|_{{}^{(k)}\mathscr{S}^+}$ with $n_3^k = 1$ the component of the outward normal of ${}^{(k)}\mathscr{S}^+$, are instead continuous (Fig. 2a).

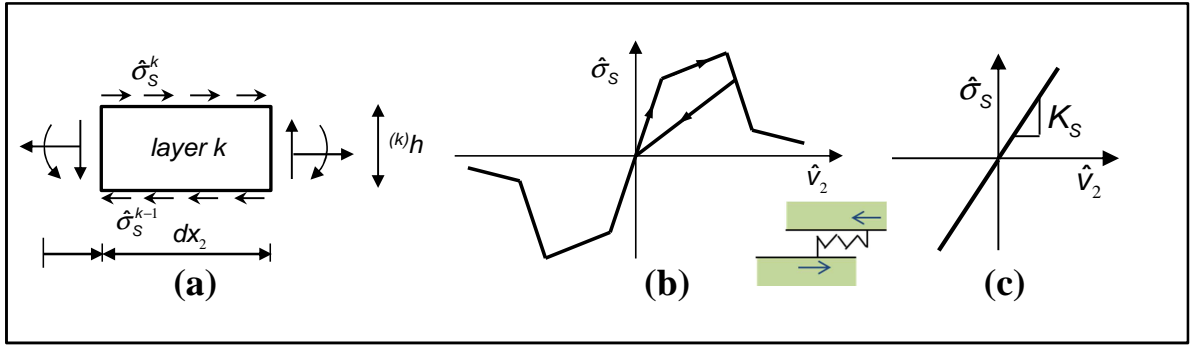

Figure 2. (a) Gross stress resultants and moments and interfacial tangential tractions acting on the infinitesimal element of layer *k*. (b) Exemplary piecewise linear cohesive traction law. (c) linear interfacial tractions law used in the explicit solutions derived in the paper, which relates interfacial shear tractions and relative sliding displacement.

The thermo-mechanical problem of Fig. 1 is solved using the multiscale approach that has been adopted in [17], for general plates, and in [18], for beams, subjected to dynamically applied mechanical loading. The approach couples a coarse-grained model and a more detailed small-scale model. The coarse-grained model, which is used to describe the global behavior of the system in [18] is a standard single-layer First Order Shear Deformation, First Order Normal Theory; for the problem at hand, where the transverse normal stresses are assumed to be negligible and the interfaces to be rigid against mode I opening, a simplified First Order Shear Deformation coarse-grained model will be used. Using higher-order coarse-grained models, as done for instance in [12], does not improve the solution in the presence of imperfect interfaces, as illustrated in [18] and remarked later in this paper. The detailed, small-scale model is a discrete-layer model which uses a classical cohesive-crack approach to treat the imperfect interfaces and delaminations.

---

[1] The assumption, which is often used in the literature, is rigorously correct only in problems where the conditions along the interfaces are purely mode II. It is acceptable in the presence of continuous interfaces, when the interfacial normal tractions are small compared to the tangential tractions and interfacial opening is prevented, e.g. by a through-thickness reinforcement [27] or other means (see [18] for a more general treatment).

According to the detailed discrete-layer approach, the interfacial tangential tractions are related to the relative sliding displacement of the layers through interfacial traction laws, which are assumed to be piece-wise linear functions, Fig. 2b, in order to approximate generally nonlinear interfacial mechanisms (e.g., perfect adhesion, bridging mechanisms produced by a through-thickness reinforcement, nails or other connectors; cohesive mechanisms arising at damaged interfaces; the presence of thin adhesive layers; elastic contact between delaminated surfaces). An arbitrary piece of the law is then defined by the affine function:

$$\hat{\sigma}_S^k(x_2) = K_S^k \hat{v}_2^k(x_2) + t_S^k \qquad (3)$$
$$\hat{v}_2^k(x_2) = B_S^k \left[ \hat{\sigma}_S^k(x_2) - t_S^k \right]$$

where $K_S^k$ and $B_S^k$ are the tangential stiffness and compliance at the interface $k$ and $t_S^k = |t_S^k| sign(\hat{v}_2^k)$ is a constant traction, with $sign(\hat{v}_2^k) = \hat{v}_2^k / |\hat{v}_2^k|$. A purely elastic interface is described by a single branch with $t_S^k = 0$ (Fig. 2c), perfectly bonded interfaces are defined by $t_S^k = 0$ and $B_S^k = 0$, which yields $\hat{v}_2^k = 0$, and fully debonded interfaces or traction-free delaminations by $t_S^k = 0$ and $K_S^k = 0$, which yields $\hat{\sigma}_S^k = 0$. For $t_S^k \neq 0$, the affine law of Eqs. (3) could describe the bridging mechanisms developed by a through-thickness reinforcement applied to a laminated composite [28], where important bridging mechanisms opposing the relative displacements develop also for small nonzero values of $\hat{v}_2^k$; for $t_S^k \neq 0$ and $K_S^k = 0$ the law could represent plastic deformations of the interlayer or through-thickness reinforcement.

**2.2 Small-scale kinematic description and downscaling relationships**

The coupling between the coarse-grained and the detailed models is performed by first assuming a two-length scale displacement field, which is described by global variables and local perturbations or enrichments. The displacement components in the layer $k$ are:

$$^{(k)}v_2(x_2, x_3) = v_{02}(x_2) + x_3 \varphi_2(x_2) + \sum_{i=1}^{k-1} \Omega_2^i(x_2)(x_3 - x_3^i) + \sum_{i=1}^{k-1} \hat{v}_2^i(x_2) \qquad (4)$$
$$^{(k)}v_3(x_2, x_3) = w_0(x_2)$$

The global variables, $v_{02}(x_2)$, $w_0(x_2)$ and $\varphi_2(x_2)$, define the displacement field of standard first order shear deformation theory, which is continuous with continuous derivatives in the thickness direction, $C_3^1$. The perturbations of the global field are assumed so that they can reproduce the local zig-zag patterns due to the multilayered structure and the jumps at the layer interfaces at $x_3 = x_3^i$ for $i=1,...,n-1$. This is done by introducing in Eq. (4) the piece-wise linear functions in $x_3$, $\Omega_2^i(x_2)(x_3 - x_3^i)$, and the jumps, $\hat{v}_2^i(x_2) = \hat{v}_2^i(x_2, x_3 = x_3^i)$, for $i=1,...,k-1$. The small-scale kinematic description of the problem is then defined by a total of 3+2×(n-1) unknown functions in the n layers.

A homogenization technique is then applied to average out the 2×(n-1) small-scale variables and obtain the macro-scale displacement field and the homogenized field equations in terms of the global variables only. Continuity conditions are first imposed on the interfacial tractions at the n-1 interfaces, which yield $^{(k)}\sigma_{23}(x_2, x_3 = x_3^k) = {}^{(k+1)}\sigma_{23}(x_2, x_3 = x_3^k)$ for $k=1,...,n-1$. Using the small-scale displacements, Eq. (4), the constitutive equations (1) and classical compatibility equations, then yields the unknown zigzag functions, $\Omega_2^k(x_2)$ for $k=1,...,n-1$, in terms of the global variables. The relationship between interfacial tractions and jumps, Eq. (3), is then used to derive the jumps, $\hat{v}_2^k(x_2)$ for $k=1,...,n-1$, in terms of the global variables. The procedure is lengthy but straightforward and is presented in details in [18].

**2.3 Macro-scale kinematic description and fields**

The zigzag functions and the jumps at the layer interfaces are substituted into Eq. (4), to obtain the macro-scale displacements in the layer $k$:

$$^{(k)}v_2(x_2, x_3) = v_{02}(x_2) + x_3\varphi_2(x_2) + \left[w_{0,2}(x_2) + \varphi_2(x_2)\right]R_{S22}^k(x_3) - \sum_{i=1}^{k-1} B_S^i t_S^i$$

$$^{(k)}v_3(x_2, x_3) = w_0(x_2)$$

(5)

where:

$$R_{S22}^k = R_{S22}^k(x_3) = \sum_{i=1}^{k-1}\left[\Lambda_{22}^{(1;i)}\left(x_3 - x_3^i\right) + \Psi_{22}^i\right]$$

$$\Lambda_{22}^{(i;j)} = {}^{(i)}C_{55}\left(\frac{1}{{}^{(j+1)}C_{55}} - \frac{1}{{}^{(j)}C_{55}}\right)$$

(6)

$$\Psi_{22}^i = {}^{(i+1)}C_{55}B_S^i\left(1+\sum_{j=1}^{i}\Lambda_{22}^{(1;j)}\right)$$

An expression that uniquely describes the displacement components in all layers can be obtained using the Heaviside's function (e.g. [18]).

The displacement field is then fully defined by the 3 global displacement variables, $v_{02}(x_2)$, $w_0(x_2)$ and $\varphi_2(x_2)$, and by parameters which account for the local enrichments and depend on the elastic constants of the material and the layup, through $\Lambda_{22}^{(i;k)}$, the geometry and the properties of the interfaces, through the $B_S^k$ and $t_S^k$ for $k=1,...,n-1$. When the reference surface coincides with the mid-surface of the bottom layer then $v_{02}(x_2) = {}^{(1)}v_2(x_2, x_3 = 0)$, and the global variables, $v_{02}$, $w_0$ and $\varphi_2$, define the displacement components of points on the reference surface and the rotations of its normal.

In the case of linear interfacial laws, Eq. (3) with $t_S^k = 0$, the equations coincide with those obtained in [14] and particularized to a shallow shell in cylindrical bending; for perfectly bonded layers, when $1/K_S^k = B_S^k = 0$ for $k=1,...,n-1$, the terms accounting for the imperfect interfaces vanish and the equations are those of classical first order zig-zag theory for fully bonded plates [8,9]; if, in addition, the material constants are continuous at the interfaces, e.g. a unidirectionally reinforced laminate, $R_{S22}^k = \Lambda_{22}^{(i;k)} = 0$ for $k=1,...,n-1$, and the Eqs. (5) describe classical first-order shear deformation single layer theory. These limits, along with the limit of fully debonded interfaces will be discussed further in Section 5.

The macro-scale strain components in the layer *k* are obtained through compatibility from the macro-scale displacement field of Eqs. (5):

$${}^{(k)}\varepsilon_{22}(x_2,x_3) = {}^{(k)}v_{2,2}(x_2,x_3) = v_{02,2}(x_2) + x_3\varphi_{2,2}(x_2) + [\varphi_{2,2}(x_2) + w_{0,22}(x_2)]R_{S22}^k(x_3) \quad (7)$$

$$2{}^{(k)}\varepsilon_{23}(x_2) = {}^{(k)}v_{2,3}(x_2,x_3) + {}^{(k)}v_{3,2}(x_2,x_3) = [w_{0,2}(x_2) + \varphi_2(x_2)]\left(1+\sum_{i=1}^{k-1}\Lambda_{22}^{(1;i)}\right) \quad (8)$$

The macro-scale stress components follow from the constitutive thermo-elastic equations (1):

$${}^{(k)}\sigma_{22}(x_2,x_3) = {}^{(k)}\bar{C}_{22}\left[v_{02,2}(x_2) + x_3\varphi_{2,2}(x_2) + [\varphi_{2,2}(x_2) + w_{0,22}(x_2)]R_{S22}^k(x_3) - {}^{(k)}\tilde{\alpha}_2\,{}^{(k)}T(x_2,x_3)\right] \quad (9)$$

$$^{(k)}\sigma_{23}(x_2) = {}^{(k)}C_{55}\left[w_{0,2}(x_2) + \varphi_2(x_2)\right]\left(1 + \sum_{i=1}^{k-1}\Lambda_{22}^{(1;i)}\right) \qquad (10)$$

The transverse shear stresses obtained from the displacement field through compatibility, Eq. (10), are constant through the thickness as a consequence of the a-priori imposition of continuity at the layer interfaces, the assumed first-order displacement field and $^{(k)}C_{55}(1+\sum_{i=1}^{k-1}\Lambda_{22}^{(1;i)}) = {}^{(k+1)}C_{55}(1+\sum_{i=1}^{k}\Lambda_{22}^{(1;i)})$ (after Eq. (6)). In addition, in systems with imperfect interfaces and delaminations, the transverse shear stresses of Eq. (10) equate the interfacial tractions and therefore decrease/vanish for decreasing/vanishing interfacial stiffness; the transverse shear strains, Eq. (8), have a similar behavior. This behavior has no detrimental effect on the equilibrium equations (see next section), and accurate predictions of transverse shear stresses and strains can be made a posteriori, $^{(k)}\sigma_{23}^{post}$ and $^{(k)}\varepsilon_{23}^{post}$, from the bending stresses of Eq. (9) by imposing local equilibrium, $^{(k)}\sigma_{22,2} + {}^{(k)}\sigma_{23,3}^{post} = 0$ and $2^{(k)}\varepsilon_{23}^{post} = 2^{(k)}\sigma_{23}^{post}/{}^{(k)}C_{55}$.

The jumps at the interfaces and the interfacial tractions in terms of the global variables, are:

$$\hat{v}_2^k(x_2) = \left[w_{0,2}(x_2) + \varphi_2(x_2)\right]\Psi_{22}^k - B_S^k t_S^k \qquad (11)$$
$$\hat{\sigma}_S^k(x_2) = K_S^k\left[w_{0,2}(x_2) + \varphi_2(x_2)\right]\Psi_{22}^k.$$

## 3. HOMOGENIZED THERMO-MECHANICAL FIELD EQUATIONS

The steady-state homogenized thermo-mechanical equilibrium equations are derived using the Principle of Virtual Work:

$$\sum_{k=1}^{n}\int_{\mathscr{S}}\int_{x_3^{k-1}}^{x_3^k}\left({}^{(k)}\sigma_{22}{}^{(k)}\delta\varepsilon_{22} + 2{}^{(k)}\sigma_{23}{}^{(k)}\delta\varepsilon_{23}\right)dx_3 dS + \sum_{k=1}^{n-1}\int_{{}^{(k)}\mathscr{S}^+}(\hat{\sigma}_S^k + t_S^k)\delta\hat{v}_2^k dS \qquad (12)$$
$$-\int_{\mathscr{S}^+}F_i^{S+}\delta v_i dS - \int_{\mathscr{S}^-}F_i^{S-}\delta v_i dS - \int_{\mathscr{B}}F_i^B \delta v_i dB = 0$$

with $i=2,3$ (a summation convention is applied to repeated subscripts). Equation (12) includes the energy contributions related to the cohesive interfaces which were erroneously neglected in all early models based on the same approach (see [17] for details). The equilibrium equations are obtained through substitution of the macro-scale strain components and displacements into Eq. (12) and then using Green's theorem whenever necessary. The equations can be stated in a form similar to that of single-layer theory to highlight similarities and differences:

$$\delta v_{02}: \quad N_{22,2} + f_2 = 0 \tag{13}$$

$$\delta \varphi_2: \quad M^b_{22,2} - Q_{2g} + f_{2m} = 0 \tag{14}$$

$$\delta w_0: \quad Q_{2g,2} + f_3 = 0 \tag{15}$$

where the force and moment resultants and loading terms are:

- normal force: $\quad N_{22}(x_2) = \sum_{k=1}^{n} \int_{x_3^{k-1}}^{x_3^k} {}^{(k)}\sigma_{22} dx_3$ (16)

- bending moment: $\quad M^b_{22}(x_2) = \sum_{k=1}^{n} \int_{x_3^{k-1}}^{x_3^k} {}^{(k)}\sigma_{22} x_3 dx_3$ (17)

- generalized transverse shear force: $\quad Q_{2g}(x_2) = Q_2^b + Q_2^z - M^z_{22,2} - M^S_{22,2} - \hat{\sigma}_2$ (18)

- transverse shear force: $\quad Q_2^b(x_2) = \sum_{k=1}^{n} \int_{x_3^{k-1}}^{x_3^k} {}^{(k)}\sigma_{23} dx_3,$ (19)

- force and moment resultants associated to the multi-layered structure:

$$Q_2^z(x_2) = \sum_{k=1}^{n} \int_{x_3^{k-1}}^{x_3^k} {}^{(k)}\sigma_{23} \sum_{i=1}^{k-1} \Lambda_{22}^{(1;i)} dx_3, \quad M^z_{22}(x_2) = \sum_{k=1}^{n} \int_{x_3^{k-1}}^{x_3^k} {}^{(k)}\sigma_{22} \sum_{i=1}^{k-1} \Lambda_{22}^{(1;i)} (x_3 - x_3^i) dx_3 \tag{20}$$

- force and moment resultants associated to the cohesive interfaces:

$$M^S_{22}(x_2) = \sum_{k=1}^{n} \int_{x_3^{k-1}}^{x_3^k} {}^{(k)}\sigma_{22} \sum_{i=1}^{k-1} \Psi^i_{22} dx_3, \quad \hat{\sigma}_2(x_2) = -\sum_{l=1}^{n-1} \left( \hat{\sigma}_S^l + t_S^l \right) \Psi^l_{22}, \tag{21}$$

- distributed tangential load: $f_2 = F_2^{S+} + F_2^{S-}$ (22)
- distributed couples: $f_{2m} = F_2^{S+} x_3^n + F_2^{S-} x_3^0 + F_2^{S+} R^n_{S22}\big|_{x_3^n}$ (23)
- distributed transverse load: $f_3 = F_3^{S+} + F_3^{S-} - F_{2,2}^{S+} R^n_{S22}\big|_{x_3^n}$ (24)

Equation (13) describes equilibrium in the longitudinal direction and coincides with the equations of classical single-layer theory (beam theory). The bending equilibrium equation, (14), which in single-layer theory would relate the derivative of the bending moment and the transverse shear force, is still valid in this model provided the shear force is substituted by a generalized transverse shear force $Q_{2g}$, given by Eq. (18). $Q_{2g}$ is statically equivalent, at any arbitrary sections of the plate with outward normal $\boldsymbol{n} = \{0,+1,0\}^T$, to the vertical equilibrant of the external forces acting on the portion of the plate to the right of the sections. In unidirectionally reinforced fully bonded systems, where $\Lambda_{22}^{(1;j)} = 0$, $t_S^k = 0$, $B_S^k = 0$ and $\hat{v}_2^k = 0$, $Q_{2g}$ equates the transverse shear force $Q_2^b$ and the equilibrium equations are those of single layer theory; in systems where the material properties are discontinuous at the interfaces and/or where the interfaces are imperfect or fully debonded, the generalized transverse shear force has the additional contributions given in Eq. (18). A generalized transverse shear stress can be introduced, $\sigma_{23g} = Q_{2g}/h$, which is the relevant internal average stress for strength predictions and averages the actual nonlinear shear stress distribution which can be obtained a posteriori by satisfying local equilibrium, $^{(k)}\sigma_{22,2} + {}^{(k)}\sigma_{23,3}^{post} = 0$, so that $\sigma_{23g} = 1/h \sum_{k=1}^{n} \int_{x_3^{k-1}}^{x_3^k} {}^{(k)}\sigma_{23}^{post} dx_3$ (as explained after Eq. (10) the shear stress obtained from compatibility is not accurate for very compliant interfaces).

The boundary conditions on $\mathscr{P}$, at $x_2 = 0, L$, with $\boldsymbol{n} = \{0,\mp 1,0\}^T$ the outward normal, are:

$$\delta v_{02}: \quad N_{22} n_2 = \tilde{N}_2^B \quad \text{or} \quad v_{02} = \tilde{v}_{02}, \tag{25}$$

$$\delta \varphi_2: \quad M_{22}^b n_2 = \tilde{M}_2^{bB} \quad \text{or} \quad \varphi_2 = \tilde{\varphi}_2, \tag{26}$$

$$\delta w_0: \quad Q_{2g} n_2 = \tilde{N}_3^B + f_{2mbc} n_2 \quad \text{or} \quad w_0 = \tilde{w}_0, \tag{27}$$

$$\delta w_{0,2}: \quad \left(M_{22}^z + M_{22}^S\right) n_2 = \tilde{M}_2^{zB} + \tilde{M}_2^{SB} \quad \text{or} \quad w_{0,2} = \tilde{w}_{0,2} \tag{28}$$

where the terms with the tilde define prescribed values of displacements, gross forces and couples and are defined in Appendix A, Eqs. (88).

Substitution of stresses and interfacial tractions into the homogenized thermo-mechanical equilibrium equations (13)-(15), using constitutive and compatibility equations and the macro-scale displacements, Eqs. (5), yields an 8th order system of ordinary differential equations:

$$C_{22}^0 v_{02,22} + \left(C_{22}^1 + C_{22}^{0S}\right) \varphi_{2,22} + C_{22}^{0S} w_{0,222} + f_2 = N_{22,2}^T \tag{29}$$

$$\left(C_{22}^1 + C_{22}^{0S}\right) v_{02,22} + \left(C_{22}^2 + 2 C_{22}^{1S} + C_{22}^{S2}\right) \varphi_{2,22} + \left(C_{22}^{1S} + C_{22}^{S2}\right) w_{0,222} \tag{30}$$
$$- \left(K^2 C_{55}^P + C_{22}^S\right) \left(\varphi_2 + w_{0,2}\right) - C_{22}^C + f_{2m} = M_{22,2}^{bT} + M_{22,2}^{zT} + M_{22,2}^{ST}$$

$$C_{22}^{0S} v_{02},_{222} + \left(C_{22}^{1S} + C_{22}^{S2}\right)\varphi_2,_{222} + C_{22}^{S2} w_0,_{2222} \qquad (31)$$
$$-\left(K^2 C_{55}^P + C_{22}^S\right)\left(\varphi_2,_2 + w_0,_{22}\right) - f_3 = M_{22}^{zT},_{22} + M_{22}^{ST},_{22}$$

The equations depend on coefficients which can be calculated a priori and describe the geometry, the layup and the status of the interfaces, $C_{22}^r, C_{22}^{rS}, C_{55}^P, C_{22}^S, C_{22}^{S2}, C_{22}^C$ for $r = 0,1,2$, Eqs. (70)-(75) of Appendix A, the applied mechanical loads, $f_2, f_{2m}, f_3$, Eqs. (22)-(24), and thermal loads, $N_{22}^T, M_{22}^{bT}, M_{22}^{zT}, M_{22}^{ST}$, Eq. (84)-(87) of Appendix A (the superscripts $z$, $S$ and $T$ indicate dependence on the multilayered structure of the material, the stiffness of the interfaces and the applied temperature, respectively).

The equations (29)-(31) also depend on a shear factor coefficient $K^2$, which has been introduced to overcome the limitations of the first order shear deformation assumption, Eqs. (5); the shear factor relates the resultant of the transverse shear stresses to the transverse shear strain through $K^2 = (Q_2^b + Q_2^z)/[C_{55}^P(\varphi_2 + w,_2)]$. Assuming $K^2 = 5/6$ in unidirectionally reinforced laminates and $K^2 = 1$ in multilayered laminates with common layups and loading conditions, yields accurate predictions of the displacements in fully bonded systems [18,8]. In systems with imperfect interfaces and delaminations, the Eqs. (29)-(31) with a constant shear factor $K^2$ under-predict the transverse displacement contribution related to the shear deformations, which progressively decreases/vanishes on decreasing/vanishing the interfacial tractions, as a consequence of the imposed continuity at the layer interfaces. In plates with continuous interfaces, the missing contribution can be fully recovered a posteriori, since the generalized shear strain associated to the generalized transverse shear force of Eq. (18) accurately describes the shear deformations in all cases, $2\varepsilon_{23g} = Q_{2g}/(K^2 C_{55}^P)$. The total transverse displacement, which accounts for bending and shear contributions, is then given by $w_0^{post}(x_2) = w_0(x_2) + w_{0add}(x_2)$, with $w_{0add}(x_2) = \int_0^{x_2}\left(Q_{2g} - Q_2^b - Q_2^z\right)/\left(K^2 C_{55}^P\right)dx_2$. When the interfaces are not continuous, e.g. plates with finite length delaminations, and in general when the stress resultants depend on the compliance of the system and the solution requires the imposition of continuity conditions between regions characterized by different transverse compliance, accurate prediction of the shear deformations might require the definition of a shear factor dependent on the interfacial stiffness.

The geometrical and mechanical boundary conditions in term of generalized displacements are given in the Eqs. (65)-(68) of Appendix A.

The system of homogenized equilibrium equations can be decoupled for efficient closed-form solution, following the procedure detailed in [29] for mechanical loading, which is based on subsequent derivations/substitutions of the Eqs. (29)-(31). The decoupling yields a $6^{th}$ order differential equation in the transverse displacement, $w_0(x_2)$, whose solution allows cascading solutions of a first order equation in $\varphi_2(x_2)$ and a $2^{nd}$ order equation in $v_{02}(x_2)$. The system of uncoupled equations is presented in the Appendix A, Eqs. (62)-(64), for general loading cases. For plates subjected to uniform transverse loading, $f_3(x_2) = q$, and/or thermal gradients generating uniform temperature distribution along $x_2$, $^{(k)}T(x_2, x_3) = {}^{(k)}T(x_3)$, the equilibrium equations assume the compact form:

$$\frac{1}{a} w_{0,222222} - w_{0,2222} + \frac{q}{d} = 0 \tag{32}$$

$$\varphi_{2,2} + w_{0,22} = \frac{1}{ac} w_{0,2222} - \frac{b_2 + b_3}{ac} q \tag{33}$$

$$v_{02,22} = \frac{C_{22}^1}{C_{22}^0} w_{0,222} - \frac{C_{22}^1 + C_{22}^{0S}}{C_{22}^0} \frac{1}{ac} w_{0,22222} \tag{34}$$

where the coefficients $a, b_2, b_3, c, d$ depend on the layup and the status of the interfaces and are given in the Eqs. (76)-(81) of Appendix A. The uncoupled system has order IX, which is higher than the order of the original system, Eqs. (29)-(31), and necessitates of the additional condition given in Eq. (69).

## 4. EXPLICIT SOLUTIONS FOR SIMPLY SUPPORTED PLATES

In this section, explicit expressions are derived for generalized displacements and stresses in simply supported wide plates subjected to uniform transverse and thermal loading, $f_3(x_2) = q$ and $^{(k)}T(x_2, x_3) = {}^{(k)}T(x_3)$, for $k = 1,...,n$ (Figs. 3b,d). The expressions have been obtained through the solution of Eqs. (32)-(34), and describe plates with an arbitrary number of layers and interfaces or delaminations. The interfaces are described by the linear traction law, $\hat{\sigma}_S^k = K_S^k \hat{v}_2^k$ for $k = 1,...,n-1$, assuming $t_S^k = 0$ in Eq. (3) (Fig. 2c); the law defines also the limit cases of perfect bonding, for $B_S^k = 1/K_S^k = 0$, $k = 1,...,n-1$, and delamination at layer $k$, for $K_S^k = 0$. The origin of the coordinate system is placed at the left plate edge and the reference surface is left arbitrary. The boundary conditions are given by Eqs. (65)-(68), with $\tilde{w}_0, \tilde{M}_2^{bB}, (\tilde{M}_2^{zB} + \tilde{M}_2^{SB}) = 0$ at $x_2 = 0, L$, $\tilde{v}_{02} = 0$ at $x_2 = 0$ and $\tilde{N}_2^B = 0$ at $x_2 = L$, along with Eq. (69).

Explicit solutions can be easily obtained for other loadings and boundary conditions using the uncoupled system Eqs. (62)-(64). In Appendix B, Eqs. (89)-(91) give closed-form solutions for plates subjected to sinusoidal transverse loads; the expressions will be used to verify the proposed model against exact 2D solutions in Section 6.

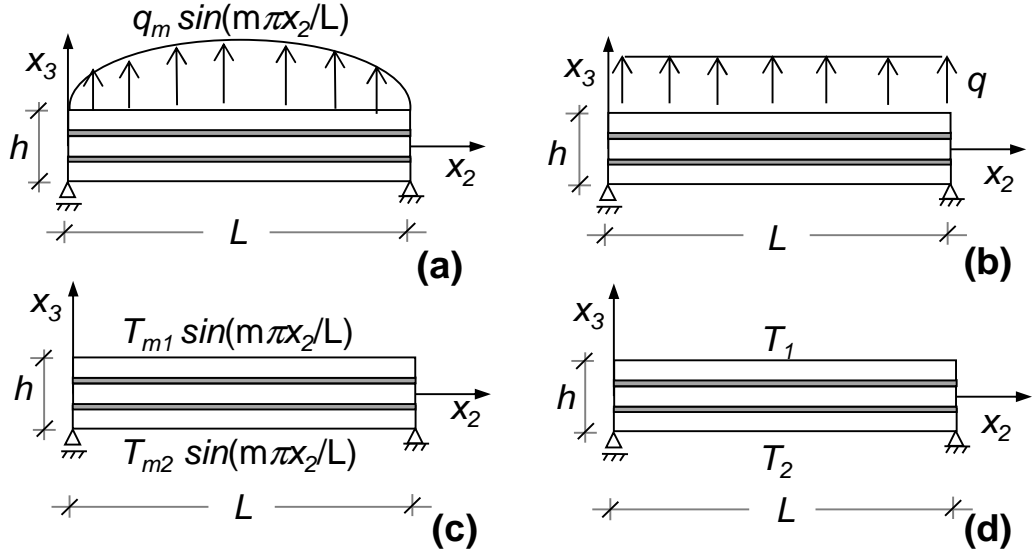

Figure 3. Boundary and loading conditions used for the explicit closed-form solutions. (a,b) Wide plate/beam subjected to mechanical transverse load; (c,d) wide plate/beam subjected to applied temperature.

### 4.1 Uniform transverse load

For uniformly distributed transverse load, $f_3(x_2) = q$, Fig. (3b), the stress resultants in the plate are $N_{22} = 0$, $M_{22}^b = q(x_2 L/2 - x_2^2/2)$ and $Q_{2g} = q(L/2 - x_2)$ from global equilibrium. The generalized displacements are:

(35)

$$w_0(x_2) + \widehat{w_{0add}}(x_2) = \frac{qL^4}{24d}\left\{\frac{x_2}{L}\left(1 - \frac{x_2}{L}\right)\left(1 + \frac{x_2}{L} - \left(\frac{x_2}{L}\right)^2\right)\right\}$$

$$+ \frac{qL^2}{d}\frac{(b_2 + b_3)d - 1}{a}\left\{\frac{\cosh\left(\sqrt{\frac{aL^2}{4}}\left(\frac{2x_2}{L} - 1\right)\right)}{aL^2\cosh\left(\sqrt{\frac{aL^2}{4}}\right)} - \frac{1}{aL^2} + \frac{1}{2}\frac{x_2}{L}\left(1 - \frac{x_2}{L}\right)\right\}\left(1 - \widehat{\frac{1}{c}}\right)$$

$$+ \widehat{\frac{qL^2}{2K^2 C_{55}^P}\frac{x_2}{L}\left(1 - \frac{x_2}{L}\right)}$$

$$\varphi_2(x_2) = \frac{qL^3}{24d}\left[\left(1 + 2\frac{x_2}{L} - 2\left(\frac{x_2}{L}\right)^2\right)\left(\frac{2x_2}{L} - 1\right)\right] +$$

$$- \frac{qL^3}{2d}\left[\left(1 - \frac{1}{c}\right)\frac{(b_2 + b_3)d - 1}{aL^2}\left(\frac{\sinh\left(\sqrt{\frac{aL^2}{4}}\left(\frac{2x_2}{L} - 1\right)\right)}{\sqrt{\frac{aL^2}{4}}\cosh\left(\sqrt{\frac{aL^2}{4}}\right)} - \frac{2x_2}{L} + 1\right)\right]$$

(36)

$$v_{02}(x_2) = -\frac{qL^3}{6d}\left[\frac{C_{22}^1}{C_{22}^0}\left(\frac{x_2}{L}\right)^2\left(\frac{3}{2} - \frac{x_2}{L}\right)\right] +$$

$$- \frac{qL^3}{d}\left[\frac{(b_2 + b_3)d - 1}{aL^2}\left(\frac{C_{22}^1}{C_{22}^0} - \frac{C_{22}^1 + C_{22}^{0S}}{cC_{22}^0}\right)\left(\frac{x_2}{L} - \frac{1}{\sqrt{a}L}\left(\frac{\sinh\left(\sqrt{\frac{aL^2}{4}}\left(\frac{2x_2}{L} - 1\right)\right)}{\cosh\left(\sqrt{\frac{aL^2}{4}}\right)} + \tanh\left(\sqrt{\frac{aL^2}{4}}\right)\right)\right)\right]$$

(37)

where the coefficients are given in Appendix A, Eqs. (70)-(83). The term $\widehat{w_{0add}}$ in the formula for the transverse displacements (35) defines the terms with the curved line on top on the right hand side, which are a-posteriori corrections to the solution of Eq. (32)-(34) to account for the full effects of the shear deformations in thick plates with very compliant interfaces or delaminations, as explained in Section 3. The importance of this term for thick anisotropic plates with very compliant interfaces has been illustrated in [18].

The bending and transverse shear stresses are obtained substituting the generalized displacements into:

$$^{(k)}\sigma_{22} = {}^{(k)}\bar{C}_{22}\left\{v_{02,2} + \varphi_{2,2} x_3 + (\varphi_{2,2} + w_{0,22})\sum_{i=1}^{k-1}\left[\Lambda_{22}^{(1;i)}(x_3 - x_3^i) + {}^{(i+1)}C_{55}B_S^i\left(1 + \sum_{j=1}^{i}\Lambda_{22}^{(1;j)}\right)\right]\right\}$$ (38)

$$\sigma_{23g} = Q_{2g}/h = (qL/2 - qx_2)/h \quad \text{(average over thickness)} \tag{39}$$

$$^{(k)}\sigma_{23}^{post}(x_3) = \int_{x_3^{k-1}}^{x_3} -{}^{(k)}\sigma_{22,2}\, dx_3 + {}^{(k-1)}\sigma_{23}^{post}(x_3 = x_3^{k-1}) \quad \text{(a posteriori from equilibrium)} \tag{40}$$

The macro-scale longitudinal displacements in the layers and the jumps at the layer interfaces are obtained substituting the global displacements, $w_0, \varphi_2$ and $v_{02}$ of Eqs. (35)-(37), into Eqs. (5) and (11) (the term $\widehat{w_{0add}}$ should not be included).

### 4.2 Uniform thermal gradient

For a uniform thermal gradient with $^{(k)}T(x_2, x_3) = {}^{(k)}T(x_3)$ for $k = 1..n-1$, the stress resultants in the plate are $N_{22} = M_{22}^b = Q_{2g} = 0$ from global equilibrium. The macro-scale displacements depend on the thermal load coefficients given in Appendix A, Eqs. (84)-(87), which can calculated once the temperature distribution has been defined from the solution of the heat conduction problem (Appendix D). The generalized displacements are:

$$w_0(x_2) = \frac{c\left((\cosh(\sqrt{a}\,x_2) - 1)\sinh(\sqrt{a}\,L) - (\cosh(\sqrt{a}\,L) - 1)\sinh(\sqrt{a}\,x_2)\right)}{C_{22}^0 d\left(K^2 C_{55}^P + C_{22}^S\right)\sinh(\sqrt{a}\,L)}$$
$$\left[(C_{22}^1 C_{22}^{1S} - C_{22}^2 C_{22}^{0S})N_{22}^T - C_{22}^0 d\left((c-1)M_{22}^{bT} - (M_{22}^{zT} + M_{22}^{ST})\right)\right]$$
$$+ \frac{(x_2 - L)x_2}{2d}\left[\frac{C_{22}^1}{C_{22}^0}N_{22}^T - M_{22}^{bT}\right] \tag{41}$$

$$\varphi_2(x_2) = \frac{\sqrt{a}(c-1)\left((\cosh(\sqrt{a}\,L) - 1)\cosh(\sqrt{a}\,x_2) - \sinh(\sqrt{a}\,x_2)\sinh(\sqrt{a}\,L)\right)}{C_{22}^0 d\left(K^2 C_{55}^P + C_{22}^S\right)\sinh(\sqrt{a}\,L)}$$
$$\left[(C_{22}^1 C_{22}^{1S} - C_{22}^2 C_{22}^{0S})N_{22}^T - C_{22}^0 d\left((c-1)M_{22}^{bT} - (M_{22}^{zT} + M_{22}^{ST})\right)\right]$$
$$+ \frac{(L - 2x_2)}{2d}\left[\frac{C_{22}^1}{C_{22}^0}N_{22}^T - M_{22}^{bT}\right] \tag{42}$$

$$v_{02}(x_2) = \frac{\sqrt{a}\,[C_{22}^{0S} - (c-1)C_{22}^1]\left[(\cosh(\sqrt{a}\,L) - 1)(\cosh(\sqrt{a}\,x_2) - 1) - \sinh(\sqrt{a}\,L)\sinh(\sqrt{a}\,x_2)\right]}{(C_{22}^0)^2 d\left(K^2 C_{55}^P + C_{22}^S\right)\sinh(\sqrt{a}\,L)}$$
$$\left[(C_{22}^1 C_{22}^{1S} - C_{22}^2 C_{22}^{0S})N_{22}^T - C_{22}^0 d\left((c-1)M_{22}^{bT} - (M_{22}^{zT} + M_{22}^{ST})\right)\right] + \frac{x_2}{C_{22}^0 d}\left[C_{22}^2 N_{22}^T - C_{22}^1 M_{22}^{bT}\right] \tag{43}$$

The bending stresses are obtained substituting the generalized displacements into:

$${}^{(k)}\sigma_{22} = {}^{(k)}\overline{C}_{22}\left\{v_{02,2} + \varphi_{2,2}x_3 + (\varphi_{2,2} + w_{0,22})\sum_{i=1}^{k-1}\left[\Lambda_{22}^{(1;i)}(x_3 - x_3^i) + {}^{(i+1)}C_{55}B_S^i\left(1 + \sum_{j=1}^{i}\Lambda_{22}^{(1;j)}\right)\right] - {}^{(k)}\tilde{\alpha}_2 T\right\}, \quad (44)$$

the generalized shear stress averaged over the thickness vanishes $\sigma_{23g} = Q_{2g}/h = 0$, and the transverse shear stresses are obtained using Eq. (40). The longitudinal macro-scale displacements in the layers and the jumps at the layer interfaces are obtained substituting the global displacements, $w_0, \varphi_2$ and $v_{02}$, into Eqs. (5) and (11).

In the special case of a unidirectionally reinforced laminate with $n-1$ imperfect interfaces having the same interfacial stiffness and subjected to a thermal gradient with linear through-thickness distribution, $T(x_3) = T_c + (x_3 - \overline{x}_3)2T_0/h$ with $\overline{x}_3 = (x_3^0 + x_3^n)/2$ and $T(x_3 = x_3^n) = T_1 = T_c + T_0$ and $T(x_3 = x_3^0) = T_2 = T_c - T_0$ (Fig. 3d), the thermal load coefficients, Eqs. (84)-(87), become:

$$N_{22}^T = C_{22}^0 \tilde{\alpha}_2 T_c + (C_{22}^1 - C_{22}^0 \overline{x}_3)\tilde{\alpha}_2 \frac{2T_0}{h}; \quad M_{22}^{bT} = C_{22}^1 \tilde{\alpha}_2 T_c + (C_{22}^2 - C_{22}^1 \overline{x}_3)\tilde{\alpha}_2 \frac{2T_0}{h}; \quad (45)$$

$$M_{22}^{ST} = C_{22}^{0S} \tilde{\alpha}_2 T_c + (C_{22}^{1S} - C_{22}^{0S}\overline{x}_3)\tilde{\alpha}_2 \frac{2T_0}{h}; \quad M_{22}^{zT} = 0.$$

The global displacements of Eqs. (41)-(43) simplify as:

$$w_0(x_2) = (L - x_2)x_2\tilde{\alpha}_2\frac{T_0}{h} \quad (46)$$

$$\varphi_2(x_2) = -(L - 2x_2)\tilde{\alpha}_2\frac{T_0}{h} \quad (47)$$

$$v_{02}(x_2) = x_2\tilde{\alpha}_2 T_c - x_2\tilde{\alpha}_2\frac{2T_0}{h}\overline{x}_3 \quad (48)$$

Eqs. (46)-(48) show that when the layers have the same elastic and thermal properties and the interfaces have the same interfacial stiffness, the solution is unaffected by the interfacial imperfections and the displacement field given above describes also the asymptotic limits of fully bonded and fully debonded layers. The jumps at the layer interfaces, Eq. (11), vanish, since $w_{0,2}(x_2) + \varphi_2(x_2) = 0$, as all stress components, Eqs. (44) and (40). As expected the solution coincides with that of classical single layer theory.

# 5. PERTURBATION ANALYSIS OF THE HOMOGENIZED FIELD EQUATIONS AND ASYMPTOTIC LIMITS

The special case of uniform transverse load and temperature distribution along the longitudinal coordinate $x_2$, will be used in this section along with a perturbation analysis to derive the asymptotic limits of the proposed model, Eq. (32)-(34), and solutions, Eqs. (35)-(37) and (41)-(43). The relevant asymptotic limits correspond to a plate where all interfaces are fully bonded (*fully bonded limit*, with $1/K_S^k \to 0$ for $k=1,...,n-1$) and to a plate where at least one of the interfaces is fully debonded (*fully debonded limit*, with $K_S^i \to 0$ for the interface $i$).

The equilibrium equations of the model in the two limits are obtained through a perturbation analysis which investigates the problem for very small values of a parameter $\delta$, where $\delta = 1/K_S^k \ll 1$ describes a perturbation with respect to the fully bonded limit and $\delta = K_S^k \ll 1$ a perturbation with respect to the fully debonded limit. The analysis also allows to investigate the multiscale dependence of the model on the parameters and highlights some relevant features.

The coefficients of the Eqs. (32)-(34), which depend on $K_S^k$, are expanded into power series of $\delta$ and their dominant orders for $\delta = 1/K_S^k \ll 1$ and $\delta = K_S^k \ll 1$ are given in the Tables 1-3 of Appendix A. The coefficient $a$, for instance, which divides the highest order term in Eq. (32) and is given in Eq. (76) is finite and positive for interfaces with finite interfacial stiffness $K_S^k$, it vanishes ($a \to 0$ as $a = O(K_S^k)$), when at least one of the interfaces is fully debonded and $\delta = K_S^i \to 0$, and is unbounded ($a \to \infty$ or $1/a \to 0$ as $1/a = O(1/(K_S^k)^2)$), when the layers are fully bonded, $\delta = 1/K_S^k \to 0$ for $k=1,...,n-1$, and the laminate is unidirectionally reinforced. In a fully bonded multilayered plate with discontinuous elastic constants at one or more interfaces, $a$ is finite and becomes unbounded ($a \to \infty$ or $1/a \to 0$) only when the discontinuities in the elastic constants vanish and $\Lambda_{22}^{(1;k)} \to 0$ for $k=1,...,n-1$ (Eq. (6)).

The generalized displacements in the Eqs. (32)-(34) can be expanded in integral powers of $\delta$ up to the second order:

$$w_0 = \overset{0}{w_0} + \delta \overset{1}{w_0} + \delta^2 \overset{2}{w_0} + O(\delta^3) \quad (49)$$

$$\varphi_2 = \overset{0}{\varphi_2} + \delta \overset{1}{\varphi_2} + \delta^2 \overset{2}{\varphi_2} + O(\delta^3)$$

$$v_{02} = \overset{0}{v_{02}} + \delta \overset{1}{v_{02}} + \delta^2 \overset{2}{v_{02}} + O(\delta^3)$$

where the superscript $\overset{i}{(\cdot)}$ on the top of a quantity is used to indicate the order of the expansion term.

Substituting the expansions into Eqs. (32)-(34) and taking the limit as $\delta \to 0$ yield the zero-order equations of the problem; the first-order equations can then be derived by dividing the original equations in terms of the perturbed variables by $\delta$ and taking the limit; and so on for the higher-order equations. The zero and higher order boundary conditions can be obtained applying a similar approach to Eqs. (65)-(69).

### 5.1 *Fully bonded asymptotic limit in unidirectionally reinforced laminates*

In this section perturbation analysis will be used to define the fully bonded asymptotic limit of the model, $\delta = 1/K_S^k \to 0$, for $k = 1,...,n-1$ in a unidirectionally reinforced laminate. The derivation will show that the zero-order equations coincide with those of Timoshenko single layer theory. This asymptotic limit of the model has been observed before in various zigzag models published in the literature (e.g. [12, 13,14]), by zeroing the coefficients which depend on the interfacial stiffness in the original equations.

When $\delta = 1/K_S^k \to 0$, for $k = 1,...,n-1$, in a unidirectionally reinforced laminate, $1/a \to 0$, $d = O(1) = \bar{C}_{22} h^3 / 12$, $C_{22}^1 / C_{22}^0 = O(1)$ and the zero order expansions of the other coefficients in Eqs. (32)-(34) are finite and given by $\overset{0}{c} = 1$ and $[(b_2 + \overset{0}{b_3})/(ac)] = 1/(K^2 C_{55} h)$ (Table 2 of Appendix A and Eqs. (70)-(83)). The resulting zero order equations are:

$$\overset{0}{w_0}_{,2222} = \frac{q}{d} \quad (50)$$

$$\overset{0}{\varphi_2}_{,2} = -\overset{0}{w_0}_{,22} - \left[ \frac{b_2 + \overset{0}{b_3}}{ac} \right] q \quad (51)$$

$$\overset{0}{v_{02}}_{,22} = \frac{C_{22}^1}{C_{22}^0} \overset{0}{w_0}_{,222} \quad (52)$$

which coincide with the classical equations of first order single-layer theory (Timoshenko beam theory), with $d = \bar{C}_{22} h^3 / 12$ the flexural stiffness, and $[ac/(\overset{0}{b_2}+b_3)] = K^2 C_{55} h$ the shear stiffness of the beam. The associated zero-order mechanical and geometrical boundary conditions also coincide with those of Timoshenko beam theory and are given by Eqs. (65)-(67),(69) by zeroing all coefficients depending on the interfacial stiffness (superscript S), while the zero-order condition (68) becomes an identity.

The higher-order equations, which are not shown here, describe the perturbation to the zero-order model for increasing values of $\delta = 1/K_S^k \ll 1$ and provide the enrichments to the global solution due to the imperfect interfaces.

The solutions of the zero-order problem of Eqs. (50)-(52) coincides with the zero-order solutions obtained through a perturbative expansion of the exact solutions of the model, Eqs. (35)-(37) and Eqs. (41)-(43), which are presented in Appendix C for uniform transverse load, Eqs. (92)-(94), and uniform thermal load, Eqs. (46)-(48).

The zero-order equation (50) has order IV, which is lower than the order VI of the original equation (32); this indicates a singularity in the model and that singular phenomena, such as boundary layers, may be expected in the solution. Indeed boundary layers are found when the model is used to analyze plates with clamped edges. In [22], for instance, the presence of a boundary layer has been demonstrated in the transverse shear force, $Q_2^b$, of a clamped beam unidirectionally reinforced and subjected to a concentrated transverse force at the free end. For $\delta = 1/K_S^k \ll 1$, $Q_2^b$ should be constant along the beam length and equate the value of the applied concentrated force, since in this limit $Q_{2g} \to Q_2^b$ (see Eq. (18) and related comments); $Q_2^b$ is instead constant over most of the domain but shows a sudden transition to zero in a boundary layer of thickness $O(1/K_S^k)$ at the clamped edge, which is a consequence of the imposed geometrical boundary conditions, Eqs. (66)-(68). The boundary layers, however, are not found neither in the force and moment resultants, which determine equilibrium in the plate, nor in the generalized displacements in the layers.

## 5.2 *Fully debonded asymptotic limit*

In this section perturbation analysis will be used to define the fully debonded asymptotic limit of the model when $\delta = K_S^i \to 0$ for at least one interface *i*. The derivation will show that the zero-order equations describe the problem of a stack of Euler-Bernoulli beams free to slide along each other.

This asymptotic limit of the homogenized model has never been derived before in the literature and demonstrates the high efficacy of the proposed model, which is able to describe with just three displacement variables the response of such highly discontinuous structural system.

When $\delta = K_S^i \to 0$ for at least one interface $i$, the finite coefficients in the equilibrium Eqs. (32)-(34) are $a \to 0$, $b_3 \to 0$, $d = O(1)$, $C_{22}^1, C_{22}^0 = O(1)$, $1/(ac) \to 1/\overset{0}{(ac)}$, $b_2 \to \overset{0}{b_2}$; the coefficient $C_{22}^{0S}/(ac)$ in Eq. (34), is instead unbounded and its expansion is $C_{22}^{0S}/(ac) = [\overset{-1}{C_{22}^{0S}/(ac)}](1/\delta) + O(1)$ (Table 3 in Appendix A and Eqs. (70)-(81)).

The expansions of the coefficients up to the relevant orders and Eqs. (49) are then substituted into Eq. (34):

$$(\overset{0}{v_{02}} + \delta \overset{1}{v_{02}})_{,22} = \frac{C_{22}^1}{C_{22}^0}(\overset{0}{w_0} + \delta \overset{1}{w_0})_{,222} - \frac{C_{22}^1}{C_{22}^0}\left[\overset{0}{\frac{1}{ac}}\right](\overset{0}{w_0} + \delta \overset{1}{w_0})_{,22222} \tag{53}$$

$$- \frac{1}{C_{22}^0}\left[\overset{-1}{\frac{C_{22}^{0S}}{ac}}\right]\frac{1}{\delta}(\overset{0}{w_0} + \delta \overset{1}{w_0})_{,22222} + O(1)$$

and taking the limit for $\delta \to 0$, Eq. (53) yields the finiteness condition, $\overset{0}{w_0}_{,22222} = 0$, and the zero-order equilibrium equation for $\overset{0}{v_{02}}$:

$$\overset{0}{v_{02}}_{,22} = \frac{C_{22}^1}{C_{22}^0}\overset{0}{w_0}_{,222} - \frac{C_{22}^1}{C_{22}^0}\left[\overset{0}{\frac{1}{ac}}\right]\overset{0}{w_0}_{,22222} - \frac{1}{C_{22}^0}\left[\overset{-1}{\frac{C_{22}^{0S}}{ac}}\right]\overset{1}{w_0}_{,22222} \tag{54}$$

which also depends on the first-order term of the expansion of $w_0$. Eqs. (32) and (33) yields the zero- and first-order equations for $w_0$ and $\varphi_2$:

$$\overset{0}{w_0}_{,222222} = 0 \tag{55}$$

$$\overset{1}{w_0}_{,222222} = [a]\left(\overset{0}{w_0}_{,2222} - \frac{q}{d}\right) \tag{56}$$

$$\overset{0}{\varphi_2}_{,2} + \overset{0}{w_0}_{,22} = \left[\overset{0}{\frac{1}{ac}}\right]\left\{\overset{0}{w_0}_{,2222} - [\overset{0}{b_2}]q\right\} \tag{57}$$

$$\overset{1}{\varphi_2},_2 + \overset{1}{w_0},_{22} = \left[\overset{0}{\frac{1}{ac}}\right] \overset{1}{w_0},_{2222} \tag{58}$$

The perturbed boundary conditions can be obtained following the same approach; the equations involve also second-order expansions of the variables and are not shown here. In combination with the condition given by Eq. (69), they yield $\overset{0}{\varphi_2} + \overset{0}{w_0},_2 = $ const. for general situations and $\overset{0}{\varphi_2} + \overset{0}{w_0},_2 = 0$ when the ends are clamped or partially fixed and in the absence of concentrated applied couples. In this case, which includes most relevant problems, such as simply-supported plates subjected to distributed mechanical and thermal loads, cantilevered plates and clamped-clamped plates, the resulting zero-order equations of the problems for $\overset{0}{w_0}$, $\overset{0}{\varphi_2}$ and $\overset{0}{v_{02}}$, Eqs. (54), (55) and (57) modify as:

$$\overset{0}{w_0},_{2222} - [\overset{0}{b_2}] q = 0 \tag{59}$$

$$\overset{0}{\varphi_2} + \overset{0}{w_0},_2 = 0 \tag{60}$$

$$\overset{0}{v_{02}},_{22} = \frac{C_{22}^1}{C_{22}^0} \overset{0}{w_0},_{222} - \frac{1}{C_{22}^0} \left[\frac{C_{22}^{0S}}{ac}\right]^{-1} \overset{1}{w_0},_{22222} \tag{61}$$

which correspond to the equations of an Euler-Bernoulli beam of flexural stiffness, $1/[\overset{0}{b_2}]$, equivalent to that of the stack of delaminated beams free to slide over each other along the delaminated interfaces (in a unidirectionally reinforced beam with $n$ equal thickness layers and $n$-1 fully debonded interfaces, $1/[\overset{0}{b_2}] = h^3 \overline{C}_{22}/(12n^2)$). The first term on the right hand side of Eq. (61) defines the global longitudinal displacements of the equivalent beam while the second term, which depend on the first-order expansion $\overset{1}{w_0}$, is necessary in this limit to describe the small scale behavior and the jumps at the interfaces.

The solution of the zero-order problem, Eqs. (59)-(61), coincides with the zero-order solutions obtained through a perturbative expansion of the exact solutions of the model, Eqs. (35)-(37) and Eqs. (41)-(43), and are presented in Appendix C for uniform transverse load, Eqs. (95)-(99), and uniform thermal load, and Eqs.(100)-(104).

## 6. APPLICATIONS AND MODEL VERIFICATION

The validation of the model formulated in this paper against exact 2D solutions has been performed in [17,18] for plates subjected to mechanical loading; thick multilayered plates, with $L/h = 4$, highly anisotropic layups, e.g. [0,90,0] with $E_T = E_L/25$, $G_{LT} = E_L/50$, $G_{TT} = E_L/125$ and $\nu_{LT} = \nu_{TT} = 0.25$, and imperfect interfaces and delaminations, have been examined (the subscripts $L$ and $T$ indicate in-plane principal material directions). The displacements of Eqs. (35)-(37) and (89)-(91) for uniform and sinusoidal transverse loads, the related stress components, interfacial tractions and displacement jumps are very accurate even in the limiting case of fully debonded plates. The reader is referred to [17,18,22] for exemplary applications of the model to multilayered plates with different layups, layer thicknesses, interfacial parameters and boundary conditions. Applications will be presented here for the thermal case using the exact 2D solution derived in the Appendix D for verification; the exact solution will be used with the simplifying assumption of interfaces rigid against mode I opening.

A multilayered anisotropic plate with $n = 3$ layers and two continuous interfaces, simply supported at the edges and subjected to an applied temperature field acting on the upper and lower surfaces (Figs. 3c,d) is considered. The layers are transversely isotropic with elastic constants $E_L$, $E_T$, $G_{LT}$, $G_{TT}$ and $\nu_{LT}$, $\nu_{TT}$; the coefficients of thermal expansion are $\alpha_L$ and $\alpha_T$ and the thermal conductivities are $K_L$ and $K_T$ (the subscripts $L$ and $T$ indicates in-plane principal material directions); the interfaces are assumed to be in perfect thermal contact, with vanishing thermal resistance $R^k = 0$ (see Appendix D), and rigid against mode I (opening) relative displacements $\hat{v}_3^k = 0$, and a linear elastic interfacial traction law is assumed to relate interfacial shear tractions and relative sliding displacement, $\hat{\sigma}_S^k = K_S^k \hat{v}_2^k$. Results will be presented for a layup [0,90,0] with $E_T = E_L/25$, $G_{LT} = E_L/50$, $G_{TT} = E_L/125$ and $\nu_{LT} = \nu_{TT} = 0.25$, $K_T = K_L/38$ and $\alpha_T = 62\alpha_L$. The response will be investigated on varying the interfacial stiffness, $K_S^k$, between the two limiting configurations $B_S^k = 1/\mathrm{K}_S^k = 0$ (fully bonded) and $K_S^k = 0$ (fully debonded). The explicit closed-form solutions presented in Eq. (41)-(43) with coefficients given by Eqs. (70)-(83), will be used; the shear factor does not influence the solutions and is assumed $K^2 = 1$.

The diagrams in Fig. 4 refer to a plate of length $L/h = 10$, subjected to a sinusoidal thermal gradient with $T(x_3 = x_3^n) = T_1 \sin(\pi x_2/L)$ and $T(x_3 = x_3^n, x_3^0) = T_2 \sin(\pi x_2/L)$ and $T_1 = T_0$ and $T_2 = -T_0$ (Fig. 3c with $m = 1$). The solution of the heat conduction problem in the Appendix D, Eqs. (108)-(113), shows that the temperature distribution in the layers is virtually linear in $x_3$, Fig. 6, and can be approximated as $T(x_2, x_3) = 2T_0/h(x_3 - \bar{x}_3) sin(\pi x_2/L)$, with $\bar{x}_3 = (x_3^0 + x_3^n)/2$. The diagrams compare longitudinal displacements, bending and transverse shear stresses obtained with the homogenized model and the exact elasticity solutions of Appendix D. Results are presented for fully bonded and fully debonded plates and for an intermediate value of interfacial stiffness. The proposed homogenized model reproduces quite accurately the complex stress and displacement distributions in the thickness of the plate, including the zig-zag behaviors and the discontinuities in the longitudinal displacements at the imperfect interfaces (jumps). The transverse displacements (not shown) are virtually coincident with the exact results for all values of interfacial stiffness.

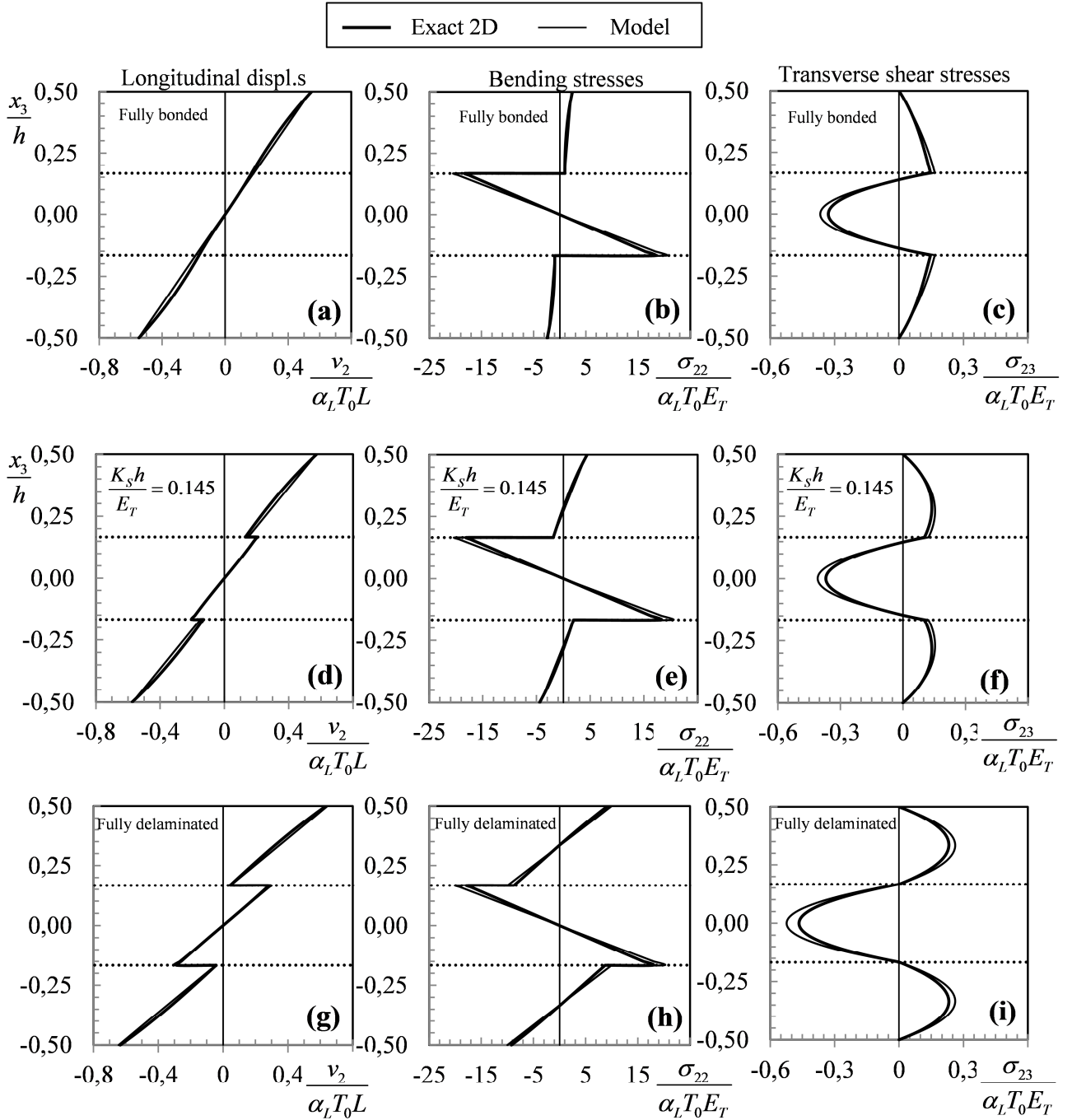

Fig. 4. Longitudinal displacements (at $x_2 = L$), transverse shear stresses (at $x_2 = 0$) and bending stresses (at $x_2 = L/2$) through thickness in a three-layer plate (0/90/0), with $L/h = 10$, subjected to a thermal gradient with $T(x_3 = x_3^n) = T_0 \sin(\pi x_2/L)$ and $T(x_3 = x_3^0) = -T_0 \sin(\pi x_2/L)$ the temperatures applied at the upper and lower surfaces, respectively (Fig. 3c). Transverse shear stresses determined from equilibrium. (a-c) Fully bonded, (g-i) fully debonded, (d-f) intermediate bonding with dimensionless interfacial stiffness $K_s h/E_T = 0.145$. Elastic constants: $E_T = E_L/25$, $G_{LT} = E_L/50$, $G_{TT} = E_L/125$ and $\nu_{LT} = \nu_{TT} = 0.25$ (as in Pagano [30]). Thin lines: proposed homogenized model; thick lines: exact 2D solution (Appendix D).

The diagrams shown in Fig. 5 refer to the geometry of the previous example with applied temperatures uniform along $x_2$, $T(x_3 = x_3^n) = T_0$ and $T(x_3 = x_3^n) = -T_0$, Fig. 3d. In this case the exact temperature distribution in the layers is piecewise linear, Eq. (110) and Fig. 6, and the model proposed here gives predictions which virtually coincide with the exact results over the entire domain but for small regions at the plate ends. In those regions, boundary layers occur in certain fields due to the imposed boundary conditions at the plate ends, Eqs. (65)-(68). The diagram in Fig. 5e shows the boundary layer in the bending stresses, which vanishes in the fully debonded asymptotic limit (the 2D solution of the problem, which has been obtained using a Fourier expansion of the solution derived in the Appendix D, is not accurate at the plate ends and is not shown).

For thick plates, e.g. $L/h = 4$, and a sinusoidally varying applied temperature, the temperature distribution in the layers become highly nonlinear in the thickness direction, Fig. 6. The exact 2D solution is strongly influenced by the through-thickness behavior and the accuracy of the model formulated in this paper, which neglects the transverse normal stresses, is reduced; the extended model formulated in [17,18], which accounts for the through thickness compressibility becomes necessary.

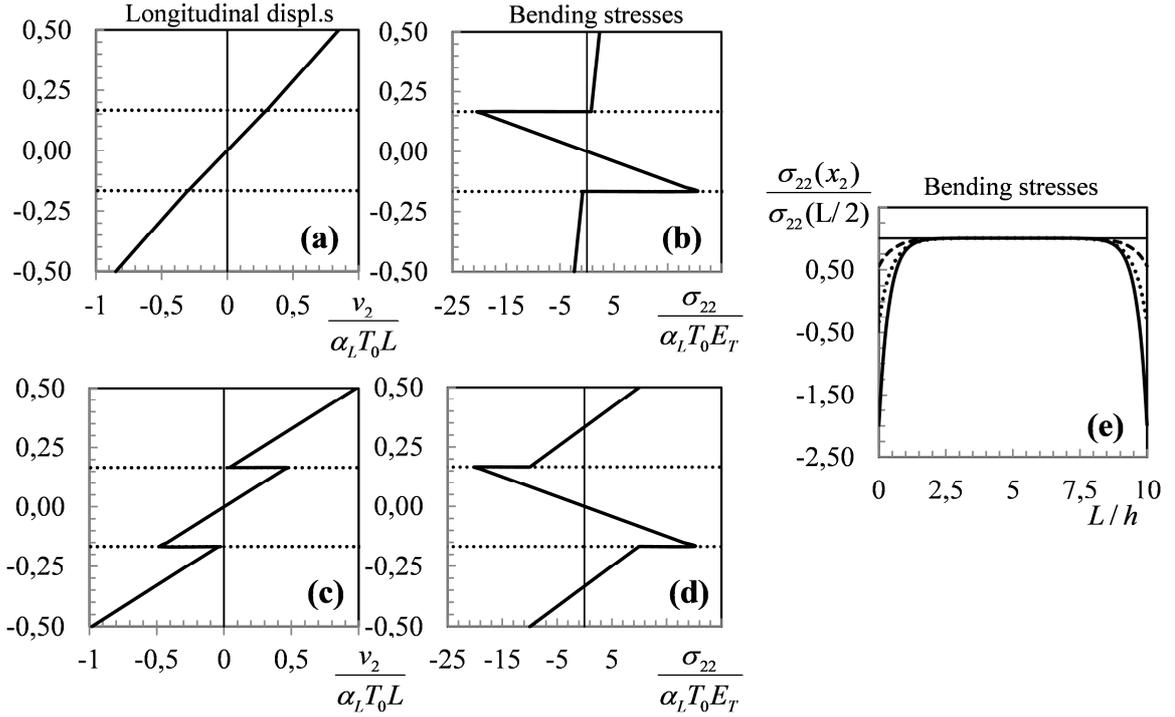

Fig. 5. Longitudinal displacements (at $x_2 = L$) and bending stresses (at $x_2 = L/2$) through thickness in a three-layer plate (0/90/0), with $L/h = 10$, subjected to a thermal gradient uniform along $x_2$ with $T(x_3 = x_3^n) = T_0$ and $T(x_3 = x_3^0) = -T_0$ the temperatures applied at the upper and lower surfaces (Fig. 3d). (a-b) Fully bonded, (c-d) fully debonded. Solutions of proposed homogenized model, Eqs. (41)-(43), virtually coincide with exact 2D solution of Appendix D. (e) Boundary layer in the bending stresses predicted by the homogenized model shown normalized to value at mid-span as function of the longitudinal coordinate, on varying the interfacial stiffness between the asymptotic limits $K_S h/E_T = \infty$ (thick), 3.65 (dot), 2.2 (dash), 0 (thin). Elastic constants: $E_T = E_L/25$, $G_{LT} = E_L/50$, $G_{TT} = E_L/125$ and $\nu_{LT} = \nu_{TT} = 0.25$ (as in Pagano [30]).

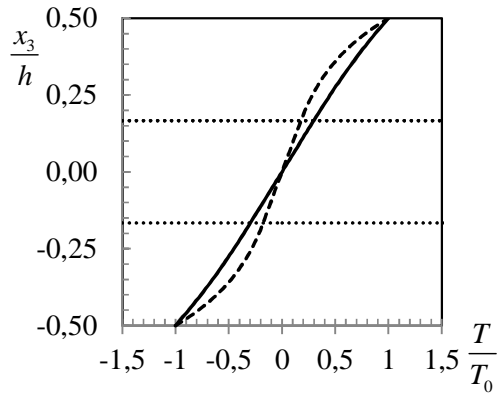

Fig. 6. Through-thickness temperature distribution in three-layer plate (0/90/0) with thermal conductivities $K_T = K_L/38$ and perfect thermal contact at the layer interfaces. The thick solid line defines the linear temperature distribution developing in a thin plate, $L/h = 10$, subjected to $T(x_2, x_3 = x_3^n, x_3^0) = T_{1,2} \sin(\pi x_2/L)$ or a thick plate, e.g. $L/h = 4$, subjected to $T(x_2, x_3 = x_3^n, x_3^0) = T_{1,2}$ with $T_1 = T_0$ and $T_2 = -T_0$. The dashed line defines the non-linear distribution which would develop in a thick plate, $L/h = 4$, subjected to the sinusoidally applied thermal gradient.

# 7. CONCLUSIONS

Explicit closed-form expressions, for efficient application in engineering practice, have been obtained for generalized displacements, stresses and interfacial jumps in multilayered wide-plates/beams subjected to thermo-mechanical loading. The expressions are general and describe plates with arbitrary layup, material constants, number and thickness of the layers and with perfect and imperfect interfaces and delaminations.

The solutions have been obtained using a multiscale model which couples a coarse-grained model (first order shear deformation theory), to describe the global fields, and a detailed small-scale model (discrete-layer cohesive-crack model), to describe the local fields. A homogenization has been applied to average out the small variables and obtain equilibrium equations and solutions in terms of the global variables, which are just three as in classical single layer theory. The model extends to thermo-mechanical loading the formulation proposed in [18] for mechanical loading.

The accuracy of the model has been verified against exact 2D solutions which have been derived in this paper for thermal loading and in [17] for mechanical loading.

The asymptotic limits of the model and the solutions have been derived by applying perturbation theory to the homogenized equilibrium equations and using a small variable $\delta \ll 1$ to describe a perturbation of the limiting problems of a plate with fully bonded interfaces ($\delta = 1/K_S^k \to 0$ with $k = 1,...,n-1$ fully bonded limit), and a plate with at least one fully debonded interface ($\delta = K_S^i \to 0$, fully debonded limit), where $K_S^k$ is the interfacial stiffness. The limiting zero-order equilibrium equations for a unidirectionally reinforced laminate coincide with those of single layer Timoshenko beam theory, in the fully bonded limit. In the fully debonded limit, the zero-order equilibrium equations of a multilayered plate coincide with those of a stack of Euler-Bernoulli beams free to slide along each other. In this second case, the first-order equations describe the local enrichments and are necessary to define the jumps in the longitudinal displacements and their effects on the small-scale local fields. The asymptotic limits/solutions and the whole transition between them is described by the three global variables of the homogenized model independently of the number of layers or imperfect interfaces and delaminations.

The perturbation technique also explains the fictitious boundary layers which may arise at the ends of the plate for certain loading and boundary conditions, e.g. clamped edges [22], and are a consequence of a singularity in the model which is introduced by the homogenization technique. The singularity however do not affect the global variables of the problem, force and moment resultants, which control equilibrium, or the generalized displacements.

Many zig-zag theories have been formulated in the literature for fully bonded systems to obtain insightful closed-form solutions and, when implemented in numerical codes, efficient solution of complex problems so overcoming the drawbacks of the computationally expensive discrete-layer models. On the other hand, for structures with imperfect interfaces and delaminations, the theory proposed in this paper for steady-state thermo-mechanical loading of wide plates and in [17,18] for dynamic loading of general plates, is the first which is based on the zig-zag original idea [8,9] and is energetically consistent and therefore accurate in describing the entire transition from fully bonded to fully debonded interfaces. The closed-form solutions of the theory offer insight into the complex behavior of multilayered plates with interlayer damage and delaminations, and the future implementation in numerical codes, following the cohesive crack approach [6], is expected to allow the efficient solution of problems dominated by progressive material damage and delamination growth.

**ACKNOWLEDGEMENTS**: support by U.S. Office of Naval Research no. N00014-14-1-0229, administered by Dr. Y.D.S. Rajapakse is gratefully acknowledged.

## APPENDIX A – Thermo-mechanical equilibrium equations

*Uncoupled homogenized equilibrium equations*

$$w_{0,222222} - aw_{0,2222} + \\
\left\{ b_1\left(f_3 + f_{2m,2} - \frac{C_{22}^1}{C_{22}^0} f_{2,2}\right) - (b_2+b_3)f_{3,22} + \left(\frac{b_2 C_{22}^1 - b_3 C_{22}^{0S}}{C_{22}^0}\right) f_{2,222} - b_2 f_{2m,222} \right\} \\
- \left\{ b_1\left(M_{22,22}^{bT} - \frac{C_{22}^1}{C_{22}^0} N_{22,22}^T\right) + b_3\left(M_{22}^{zT} + M_{22}^{ST}\right)_{,2222} \right. \\
\left. + \left(\frac{b_2 C_{22}^1 - b_3 C_{22}^{0S}}{C_{22}^0}\right) N_{22,2222}^T - b_2 M_{22,2222}^{bT} \right\} = 0 \tag{62}$$

$$\varphi_{2,2} = -w_{0,22} + \frac{1}{ac} w_{0,2222} - \frac{1}{ac}\left\{ (b_2+b_3)f_3 - \left(\frac{b_2 C_{22}^1 - b_3 C_{22}^{0S}}{C_{22}^0}\right) f_{2,2} + b_2 f_{2m,2} \right\} + \\
+ \frac{1}{ac}\left\{ -b_3\left(M_{22,2}^{zT} + M_{22,2}^{ST}\right) - \left(\frac{b_2 C_{22}^1 - b_3 C_{22}^{0S}}{C_{22}^0}\right) N_{22,2}^T + b_2 M_{22,2}^{bT} \right\} \tag{63}$$

$$v_{02,22} = \frac{C_{22}^1}{C_{22}^0} w_{0,222} - \frac{C_{22}^1 + C_{22}^{0S}}{C_{22}^0}\frac{1}{ac} w_{0,22222} - \frac{1}{C_{22}^0} f_2 \\
+ \frac{C_{22}^1 + C_{22}^{0S}}{C_{22}^0}\frac{1}{ac}\left\{ (b_2+b_3)f_{3,2} - \left(\frac{b_2 C_{22}^1 - b_3 C_{22}^{0S}}{C_{22}^0}\right) f_{2,22} + b_2 f_{2m,22} \right\} + \frac{1}{C_{22}^0} N_{22,2}^T \\
- \frac{C_{22}^1 + C_{22}^{0S}}{C_{22}^0}\frac{1}{ac}\left\{ -b_3\left(M_{22,22}^{zT} + M_{22,22}^{ST}\right) - \left(\frac{b_2 C_{22}^1 - b_3 C_{22}^{0S}}{C_{22}^0}\right) N_{22,22}^T + b_2 M_{22,22}^{bT} \right\} \tag{64}$$

*Geometrical and Mechanical Boundary Conditions*

$$\delta v_{02}: \quad \left[ v_{02,2} C_{22}^0 + \varphi_{2,2}\left(C_{22}^1 + C_{22}^{0S}\right) + w_{0,22} C_{22}^{0S} - N_{22}^T \right] n_2 = \tilde{N}_2^B \quad \text{or} \tag{65}$$

$$v_{02} = \tilde{v}_{02}$$

$\delta w_0 :$ (66)

$$\left\{ \begin{array}{l} -v_{02,22}C_{22}^{0S} - \varphi_{2,22}\left(C_{22}^{1S} + C_{22}^{S2}\right) + \left(w_{0,2} + \varphi_2\right)\left[K^2 C_{55}^P + C_{22}^S\right] \\ -w_{0,222}C_{22}^{S2} + C_{22}^C + \left(M_{22,2}^{zT} + M_{22,2}^{ST}\right) - f_{2mbc} \end{array} \right\} n_2 = \tilde{N}_3^B \quad \text{or} \quad w_0 = \tilde{w}_0$$

$\delta\varphi_2 :\ \left[v_{02,2}C_{22}^1 + \varphi_{2,2}\left(C_{22}^2 + C_{22}^{1S}\right) + w_{0,22}C_{22}^{1S} - M_{22}^{bT}\right]n_2 = \tilde{M}_2^{bB}$ or $\varphi_2 = \tilde{\varphi}_2$ (67)

$\delta w_{0,2} :\ \left[v_{02,2}C_{22}^{0S} + \varphi_{2,2}\left(C_{22}^{1S} + C_{22}^{S2}\right) + w_{0,22}C_{22}^{S2} - \left(M_{22}^{zT} + M_{22}^{ST}\right)\right]n_2 = \tilde{M}_2^{zB} + \tilde{M}_2^{SB}$ or $w_{0,2} = \tilde{w}_{0,2}$ (68)

Additional condition needed to define the constants of integration when using the uncoupled Eqs. (62)-(64):

$$\left(C_{22}^1 + C_{22}^{0S}\right)v_{02,22} + \left(C_{22}^2 + 2C_{22}^{1S} + C_{22}^{S2}\right)\varphi_{2,22} + \left(C_{22}^{1S} + C_{22}^{S2}\right)w_{0,222} + \quad (69)$$
$$-\left(K^2 C_{55}^P + C_{22}^S\right)\left(\varphi_2 + w_{0,2}\right) - C_{22}^C + f_{2m} = M_{22,2}^{bT} + M_{22,2}^{zT} + M_{22,2}^{ST}$$

*Coefficients: geometry, layup, status of the interfaces*

$$C_{22}^r = \sum_{k=1}^n {}^{(k)}\overline{C}_{22} \int_{x_3^{k-1}}^{x_3^k} (x_3)^r dx_3, \quad (70)$$

$$C_{22}^{rS} = \sum_{k=1}^n {}^{(k)}\overline{C}_{22} \int_{x_3^{k-1}}^{x_3^k} (x_3)^r R_{S22}^k dx_3, \quad (71)$$

$$C_{22}^{S2} = \sum_{k=1}^n {}^{(k)}\overline{C}_{22} \int_{x_3^{k-1}}^{x_3^k} \left(R_{S22}^k\right)^2 dx_3 \quad (72)$$

$$C_{22}^S = \sum_{k=1}^{n-1} \left(\Psi_{22}^k\right)^2 / B_S^k \quad (73)$$

$$C_{55}^P = \sum_{k=1}^n {}^{(k)}C_{55} \int_{x_3^{k-1}}^{x_3^k} \left(1 + \sum_{i=1}^{k-1} \Lambda_{22}^{(1;i)}\right)^2 dx_3 \quad (74)$$

$$C_{22}^C = \sum_{k=1}^{n-1} t_S^{(k)} \Psi_{22}^{(k)} \quad (75)$$

$$a = \frac{\left(K^2 C_{55}^P + C_{22}^S\right)\left[C_{22}^2 C_{22}^0 - \left(C_{22}^1\right)^2\right]C_{22}^0}{\left[C_{22}^{S2}C_{22}^0 - \left(C_{22}^{0S}\right)^2\right]\left[C_{22}^2 C_{22}^0 - \left(C_{22}^1\right)^2\right] - \left(C_{22}^0 C_{22}^{1S} - C_{22}^1 C_{22}^{0S}\right)^2} \quad (76)$$

$$b_1 = \frac{\left(K^2 C_{55}^P + C_{22}^S\right)\left(C_{22}^0\right)^2}{\left[C_{22}^{S2} C_{22}^0 - \left(C_{22}^0\right)^2\right]\left[C_{22}^2 C_{22}^0 - \left(C_{22}^1\right)^2\right] - \left(C_{22}^0 C_{22}^{1S} - C_{22}^1 C_{22}^{0S}\right)^2} \tag{77}$$

$$b_2 = \frac{\left\{\left[C_{22}^{S2} C_{22}^0 - \left(C_{22}^{0S}\right)^2\right] + \left(C_{22}^0 C_{22}^{1S} - C_{22}^1 C_{22}^{0S}\right)\right\} C_{22}^0}{\left[C_{22}^{S2} C_{22}^0 - \left(C_{22}^{0S}\right)^2\right]\left[C_{22}^2 C_{22}^0 - \left(C_{22}^1\right)^2\right] - \left(C_{22}^0 C_{22}^{1S} - C_{22}^1 C_{22}^{0S}\right)^2} \tag{78}$$

$$b_3 = \frac{\left\{\left(C_{22}^0 C_{22}^{1S} - C_{22}^1 C_{22}^{0S}\right) + \left[C_{22}^2 C_{22}^0 - \left(C_{22}^1\right)^2\right]\right\} C_{22}^0}{\left[C_{22}^{S2} C_{22}^0 - \left(C_{22}^{0S}\right)^2\right]\left[C_{22}^2 C_{22}^0 - \left(C_{22}^1\right)^2\right] - \left(C_{22}^0 C_{22}^{1S} - C_{22}^1 C_{22}^{0S}\right)^2} \tag{79}$$

$$c = \frac{\left(C_{22}^0 C_{22}^{1S} - C_{22}^1 C_{22}^{0S}\right) + \left[C_{22}^2 C_{22}^0 - \left(C_{22}^1\right)^2\right]}{\left[C_{22}^2 C_{22}^0 - \left(C_{22}^1\right)^2\right]} \tag{80}$$

$$d = \frac{a}{b_1} = \frac{\left[C_{22}^2 C_{22}^0 - \left(C_{22}^1\right)^2\right]}{C_{22}^0} \tag{81}$$

$$\frac{1}{ac} = \frac{\left[C_{22}^{S2} C_{22}^0 - \left(C_{22}^{0S}\right)^2\right]\left[C_{22}^2 C_{22}^0 - \left(C_{22}^1\right)^2\right] - \left(C_{22}^0 C_{22}^{1S} - C_{22}^1 C_{22}^{0S}\right)^2}{\left(K^2 C_{55}^P + C_{22}^S\right) C_{22}^0 \left\{\left(C_{22}^0 C_{22}^{1S} - C_{22}^1 C_{22}^{0S}\right) + \left[C_{22}^2 C_{22}^0 - \left(C_{22}^1\right)^2\right]\right\}} \tag{82}$$

$$\frac{b_2 + b_3}{a} = \frac{\left[C_{22}^{S2} C_{22}^0 - \left(C_{22}^{0S}\right)^2\right] + 2\left(C_{22}^0 C_{22}^{1S} - C_{22}^1 C_{22}^{0S}\right) + \left[C_{22}^2 C_{22}^0 - \left(C_{22}^1\right)^2\right]}{\left(K^2 C_{55}^P + C_{22}^S\right)\left(C_{22}^2 C_{22}^0 - (C_{22}^1)^2\right)} \tag{83}$$

*Coefficients: thermal loads*

$$N_{22}^T = \sum_{k=1}^n {}^{(k)}\overline{C}_{22} \int_{x_3^{k-1}}^{x_3^k} {}^{(k)}\tilde{\varepsilon}_{22}^t dx_3, \tag{84}$$

$$M_{22}^{bT} = \sum_{k=1}^n {}^{(k)}\overline{C}_{22} \int_{x_3^{k-1}}^{x_3^k} {}^{(k)}\tilde{\varepsilon}_{22}^t x_3 dx_3 \tag{85}$$

$$M_{22}^{ST} = \sum_{k=1}^n {}^{(k)}\overline{C}_{22} \int_{x_3^{k-1}}^{x_3^k} {}^{(k)}\tilde{\varepsilon}_{22}^t \sum_{i=1}^{k-1} \Psi_{22}^i dx_3, \tag{86}$$

$$M_{22}^{zT} = \sum_{k=1}^n {}^{(k)}\overline{C}_{22} \int_{x_3^{k-1}}^{x_3^k} {}^{(k)}\tilde{\varepsilon}_{22}^t \sum_{i=1}^{k-1} \Lambda_{22}^{(1;i)}\left(x_3 - x_3^i\right) dx_3 \tag{87}$$

*Coefficients: prescribed forces and moments at the plate ends*

$$\tilde{N}_i^B = \sum_{k=1}^{n} \int_{x_3^{k-1}}^{x_3^k} {}^{(k)}F_i^B dx_3, \text{ for } i = 2,3 \qquad \tilde{M}_2^{bB} = \sum_{k=1}^{n} \int_{x_3^{k-1}}^{x_3^k} {}^{(k)}F_2^B x_3 dx_3,$$

$$\tilde{M}_2^{SB} = \sum_{k=1}^{n} \int_{x_3^{k-1}}^{x_3^k} {}^{(k)}F_2^B \sum_{i=1}^{k-1} \Psi_{22}^i dx_3, \qquad \tilde{M}_2^{zB} = \sum_{k=1}^{n} \int_{x_3^{k-1}}^{x_3^k} {}^{(k)}F_2^B \sum_{i=1}^{k-1} \Lambda_{22}^{(1;i)} \left(x_3 - x_3^i\right) dx_3,$$

$$f_{2mbc} = F_2^{S+} R_{S22}^n \big|_{x_3^n}$$

(88)

*Order of the coefficients in the asymptotic limits of fully bonded and fully debonded plates*

The coefficients of the differential equations (62)-(64) and (32)-(34) can be expanded into power series of $\delta$. The tables below defines the dominant orders of the coefficients in the two asymptotic limits.

Table 1 – Fully bonded multilayered plate, $\delta = 1/K_S^k \ll 1$ for $k = 1,...,n-1$:

| Vanishing coefficients | Finite value coefficients | Unbounded coefficients |
|---|---|---|
| $B_S^k = 1/K_S^k$ <br> $O(1/K_S^k): \Psi_{22}^k, C_{22}^S;$ | $O(1): C_{22}^0, C_{22}^1, C_{22}^2, C_{55}^P; a,c,b_1,b_2,b_3,d,$ <br> $C_{22}^{0S}, C_{22}^{1S}, C_{22}^{S2}, R_{S22}^k, C_{22}^C = 0;$ | $K_S^k \to \infty;$ |

Table 2 – Fully bonded unidirectionally reinforced plate, $\delta = 1/K_S^k \ll 1$ for $k = 1,...,n-1$:

| Vanishing coefficients | Finite value coefficients | Unbounded coefficients |
|---|---|---|
| $B_S^k = 1/K_S^k$ <br> $O(1/(K_S^k)^2): 1/a, 1/(ac), C_{22}^{S2};$ <br> $O(1/K_S^k): \Psi_{22}^k, R_{S22}^k, C_{22}^{0S}, C_{22}^{1S}, C_{22}^S;$ | $O(1): C_{22}^0, C_{22}^1, C_{22}^2, C_{55}^P,$ <br> $c \to 1, d = \dfrac{\overline{C}_{22} h^3}{12},$ <br> $\dfrac{b_2 + b_3}{a} \to \dfrac{1}{K^2 C_{55} h}, C_{22}^C = 0$ | $K_S^k \to \infty;$ <br> $O((K_S^k)^2): a, b_1, b_3;$ <br> $O(K_S^k): b_2;$ |

Table 3 – Fully debonded plate, $\delta = K_S^i \ll 1$ for at least one interface $i$:

| Vanishing coefficients | Finite value coefficients | Unbounded coefficients |
|---|---|---|
| $K_S^k$ <br> $O(K_S^k): a, b_1, b_3;$ | $O(1): C_{22}^0, C_{22}^1, C_{22}^2, C_{55}^P, b_2^{(*)},$ <br> $ac, d, C_{22}^{0S}/c, C_{22}^C = 0;$ | $B_S^k \to \infty;$ <br> $O(1/K_S^k): \Psi_{22}^k, R_{S22}^k, c, C_{22}^{0S}, C_{22}^{1S}, C_{22}^S;$ <br> $O(1/(K_S^k)^2): C_{22}^{S2};$ |

${}^{(*)} \overset{0}{b}_2 = \dfrac{12}{\overline{C}_{22} h^3} n^2$, for unidirectionally reinforced laminates with equal thickness layers

# APPENDIX B – Explicit solution for simply supported plate subjected to sinusoidal transverse load

In a simply supported plate subjected to a sinusoidal transverse load, $f_3(x_2) = q_m \sin(px_2)$, with $p = m\pi/L$ and $m \in \mathbb{N}$, Fig. (3a), the generalized displacements are:

$$w_0(x_2) + \widehat{w_{0add}} = \frac{q_m}{p^2}\left\{\left(\frac{1}{p^2 d} + \frac{(b_2 + b_3)}{a}\right)\left(1 + \frac{p^2}{a}\right)^{-1}\right.$$

$$\left.+\overbrace{\left[\frac{b_1}{a^2 c} - \frac{(b_2 + b_3)}{ac}\right]\left(1 + \frac{p^2}{a}\right)^{-1}} + \frac{1}{K^2 C_{55}^P}\right\}\sin(px_2) \tag{89}$$

$$\varphi_2(x_2) = -\frac{q_m}{p^3 d}\left\{1 + \left[1 + (c-1)(b_2 + b_3)d\right]\frac{p^2}{ac}\right\}\left(1 + \frac{p^2}{a}\right)^{-1}\cos(px_2) \tag{90}$$

$$v_{02}(x_2) = \frac{q_m}{p^3 d}\left\{\frac{C_{22}^1}{C_{22}^0} + \left[\frac{C_{22}^1 + C_{22}^{0S} + (cC_{22}^1 - C_{22}^1 - C_{22}^{0S})(b_2 + b_3)d}{C_{22}^0}\right]\frac{p^2}{ac}\right\}$$

$$\left(1 + \frac{p^2}{a}\right)^{-1}\left[\cos(px_2) - 1\right] \tag{91}$$

# APPENDIX C – Asymptotic limits of the macro-scale generalized displacements in simply supported plates

## C.1 Uniform transverse load – fully bonded limit

When $\delta = 1/K_S^k \to 0$ for $k = 1,...,n-1$, $1/a \to 0$, $d = O(1) = \bar{C}_{22}h^3/12$, $C_{22}^1/C_{22}^0 = O(1)$ and the zero order expansions of the other coefficients in the solutions (35)-(37) are finite and given by $\overset{0}{c} = 1$ and $[\overset{0}{(b_2 + b_3)}/(ac)] = 1/(K^2 C_{55}h)$. In the zero-order expansion in integral powers of $\delta = 1/K_S^k$ of Eqs. (35)-(37), the terms with the hyperbolic functions vanish and the two terms with the curved line on top cancel each other, $\widehat{w_{0add}} = 0$. The zero-order expansion define the global displacements, which coincide with the solution of classical first order shear deformation single-layer theory. The resulting zero-order displacements are:

$$\overset{0}{w_0}(x_2) = \frac{qL^4}{2\bar{C}_{22}h^3}\left[\frac{x_2}{L}\left(1-\frac{x_2}{L}\right)\left(1+\frac{x_2}{L}-\left(\frac{x_2}{L}\right)^2\right)\right] + \frac{qL^2}{2K^2C_{55}h}\frac{x_2}{L}\left(1-\frac{x_2}{L}\right) \tag{92}$$

$$\overset{0}{\varphi_2}(x_2) = \frac{qL^3}{2\bar{C}_{22}h^3}\left[\left(1+2\frac{x_2}{L}-2\left(\frac{x_2}{L}\right)^2\right)\left(\frac{2x_2}{L}-1\right)\right] \tag{93}$$

$$\overset{0}{v_{02}}(x_2) = \frac{qL^3}{\bar{C}_{22}h^3}\frac{C_{22}^1}{C_{22}^0}\left(\frac{x_2}{L}\right)^2\left(\frac{2x_2}{L}-3\right) \tag{94}$$

The maximum deflection at $x_2 = L/2$ is $\overset{0}{w_0}(x_2 = L/2) = \frac{5qL^4}{32\bar{C}_{22}h^3} + \frac{qL^2}{8K^2C_{55}h}$ with $\bar{C}_{22}$ the reduced longitudinal stiffness, Eq. (1).

### C.2 Uniform transverse load – fully debonded limit

When $\delta = K_S^i \to 0$ for at least one interface $i$, the finite coefficients in the solutions (35)-(37) are $a \to 0$, $b_3 \to 0$, $d = O(1)$, $C_{22}^1, C_{22}^0 = O(1)$, $1/(ac) \to 1/\overset{0}{(ac)}$, $b_2 \to \overset{0}{b_2}$, $C_{22}^{0S}/c \to \left(C_{22}^{0S}/c\right)$; while $c \to \infty$ is unbounded. The zero- and first-order terms of the expansion in integral power of $\delta$ of Eqs. (35)- (37), define the global displacements and the small-scale enrichments, respectively. Up to a rigid longitudinal translation, which depend on the position of the assumed reference surface and vanishes when $x_3^0 = -{}^{(1)}h/2$, the first order expansions $w_0 + \widehat{w_{0add}} = \overset{0}{w_0} + \widehat{w_{0add}}$, $\varphi_2 = \overset{0}{\varphi_2}$ and $v_{02} = \overset{0}{v_{02}}$, define the generalized displacements of a stack of Timoshenko beams free to slide over each other: $\overset{0}{w_0}$, $\overset{0}{\varphi_2}$ and $\overset{0}{v_{02}}$ are the solutions of classical Euler-Bernoulli single-layer theory; $\widehat{w_{0add}}$ accounts for shear deformations. The first-order expansions $w_0 = \overset{0}{w_0} + \delta \overset{1}{w_0}$ and $\varphi_2 = \overset{0}{\varphi_2} + \delta \overset{1}{\varphi_2}$ are instead needed to derive the longitudinal displacements in the layers and to fully describe the small-scale behavior (jumps, stresses and displacements in the layers) using Eqs. (5), (11), (38) and (40). The expansions are:

$$\overset{0}{w_0}(x_2) + \widehat{w_{0add}}(x_2) = \frac{qL^4}{24}\overset{0}{b_2}\frac{x_2}{L}\left(1-\frac{x_2}{L}\right)\left(1+\frac{x_2}{L}-\left(\frac{x_2}{L}\right)^2\right) + \widehat{\frac{qL^2}{2}\frac{1}{K^2C_{55}^P}\frac{x_2}{L}\left(1-\frac{x_2}{L}\right)} \tag{95}$$

$$\overset{1}{w_0}(x_2)\delta = -\left[\frac{qL^6}{240}\left(\overset{0}{b_2}-\frac{1}{d}\right)\left(\frac{x_2}{L}-\frac{5}{3}\left(\frac{x_2}{L}\right)^3+\left(\frac{x_2}{L}\right)^5-\frac{1}{3}\left(\frac{x_2}{L}\right)^6\right)\right]\overset{1}{[a]}\delta \tag{96}$$

$$\overset{0}{\varphi_2}(x_2) = \frac{qL^3}{24}\overset{0}{b_2}\left(1+2\frac{x_2}{L}-2\left(\frac{x_2}{L}\right)^2\right)\left(\frac{2x_2}{L}-1\right) \tag{97}$$

$$\overset{1}{\varphi_2}(x_2)\delta = -\left(\overset{0}{b_2}-\frac{1}{d}\right)\left(\frac{2x_2}{L}-1\right)\left[\frac{qL^3}{24}\left(\frac{\overset{0}{1}}{ac}\right)\left(1+2\frac{x_2}{L}-2\left(\frac{x_2}{L}\right)^2\right)+\frac{qL^5}{240}\left(1+\frac{x_2}{L}-\left(\frac{x_2}{L}\right)^2\right)^2\right]\overset{1}{[a]}\delta \tag{98}$$

$$\overset{0}{v_{02}}(x_2) = -\frac{qL^3}{6}\left[\overset{0}{b_2}\frac{C_{22}^1}{C_{22}^0}+\left(\frac{1}{d}-\overset{0}{b_2}\right)\left(\frac{\overset{0}{C_{22}^{0S}}}{cC_{22}^0}\right)\right]\left(\frac{x_2}{L}\right)^2\left(\frac{3}{2}-\frac{x_2}{L}\right) \tag{99}$$

where $1/\overset{0}{b_2}$ is the flexural stiffness of the layer assembly ($1/\overset{0}{b_2} \to h^3\overline{C}_{22}/(12n^2)$) in a unidirectionally reinforced laminate with $n$ equal thickness fully delaminated layers). The maximum deflection at $x_2 = L/2$ is $\overset{0}{w_0}(x_2 = L/2) = \frac{5\overset{0}{b_2}qL^4}{384}+\frac{qL^2}{8K^2C_{55}h}$ which becomes $\overset{0}{w_0}(x_2 = L/2) = \frac{5qL^4}{32\overline{C}_{22}h^3}+\frac{qL^2}{8K^2C_{55}h}$ in a plate with $n$ equal thickness layers.

The need of using the first order expansions of the transverse displacement and rotation to define the small-scale behavior is made clear by looking at the asymptotic expansion of Eq. (11) $\hat{v}_2^k = \overset{0}{\hat{v}_2^k}+O(\delta) = \left[(\overset{0}{w_{0,2}}+\overset{0}{\varphi_2})+(\overset{1}{w_{0,2}}+\overset{1}{\varphi_2})\delta\right]B_S^k(1+\sum_{j=1}^{k}\Lambda_{22}^{(1;j)})^{(k+1)}C_{55}+O(\delta)$; the term $(\overset{0}{w_{0,2}}+\overset{0}{\varphi_2}) = 0$, after Eqs. (96) and (97); then, since $\delta = 1/B_S^k$, the zero order expansion of the jump is finite and given by $\overset{0}{\hat{v}_2^k} = (\overset{1}{w_{0,2}}+\overset{1}{\varphi_2})(1+\sum_{j=1}^{k}\Lambda_{22}^{(1;j)})^{(k+1)}C_{55}$.

### C.3 Uniform thermal load – fully debonded limit

When $\delta = K_S^i \to 0$ for at least one interface $i$, the finite coefficients in the solutions (41)-(43) are $a \to 0$, $d = O(1)$, $C_{22}^1, C_{22}^0 = O(1)$, $C_{22}^{0S}/c \to \left(\overset{0}{C_{22}^{0S}}/c\right)$; while $c \to \infty$ is unbounded. As explained above for the asymptotic limit of the case with uniform transverse load, the zero- and first-order terms of the expansion in integral power of $\delta$ of Eqs. (41)-(43) define the global displacements (assembly of Timoshenko layers free to slide over each other) and the small-scale enrichment, respectively. The zero order expansions are:

$$\overset{0}{w_0}(x_2) = \frac{1}{C_{22}^0 d} \frac{x_2(x_2 - L)}{2} \left\{ \left( \frac{a(C_{22}^{S2}C_{22}^1 - C_{22}^{0S}C_{22}^{1S})}{C_{22}^S} \right) \overset{0}{N_{22}^T} + \left( \frac{a\left((C_{22}^{0S})^2 - C_{22}^0 C_{22}^{S2}\right)}{C_{22}^S} \right) \overset{0}{M_{22}^{bT}} \right. \tag{100}$$

$$\left. + C_{22}^0 d \left( \frac{ac}{C_{22}^S} \overset{0}{(M_{22}^{zT} + M_{22}^{ST})} \right) \right\}$$

$$\overset{0}{\varphi_2}(x_2) = \frac{1}{C_{22}^0 d} \frac{L - 2x_2}{2} \left\{ \left( \frac{a(C_{22}^1 C_{22}^{S2} - C_{22}^{0S}C_{22}^{1S})}{C_{22}^S} \right) \overset{0}{N_{22}^T} + \left( \frac{a((C_{22}^{0S})^2 - C_{22}^0 C_{22}^{S2})}{C_{22}^S} \right) \overset{0}{M_{22}^{bT}} \right. \tag{101}$$

$$\left. + C_{22}^0 d \left( \frac{ac}{C_{22}^S} \overset{0}{(M_{22}^{zT} + M_{22}^{ST})} \right) \right\}$$

$$\overset{0}{v_{02}}(x_2) = \frac{x_2}{C_{22}^0 d} \left\{ \left( \frac{a\left(C_{22}^{S2}C_{22}^2 - (C_{22}^{1S})^2\right)}{C_{22}^S} \right) \overset{0}{N_{22}^T} + \left( \frac{a\left(C_{22}^{0S}C_{22}^{1S} - C_{22}^{S2}C_{22}^1\right)}{C_{22}^S} \right) \overset{0}{M_{22}^{bT}} \right. \tag{102}$$

$$\left. + \left( \frac{a\left(cC_{22}^1 - C_{22}^{0S}\right)}{C_{22}^S} \overset{0}{(M_{22}^{zT} + M_{22}^{ST})} \right) \right\}$$

And the first order expansions:

$$\overset{1}{w_0}(x_2) = \frac{1}{C_{22}^0 d} \frac{x_2(x_2 - L)}{24} \left\{ \left( \frac{ac(C_{22}^{S0}C_{22}^2 - C_{22}^{1S}C_{22}^1)}{C_{22}^S} \left(L^2 + x_2 L - x_2^2\right) + \frac{12K^2 C_{55}^P C_{22}^1}{aC_{22}^S} \right) \overset{0}{N_{22}^T} \right. \tag{103}$$

$$+ \left( \frac{ac^2 d C_{22}^0}{C_{22}^S} \left(L^2 + x_2 L - x_2^2\right) + \frac{12 C_{22}^0 \left(K^2 C_{55}^P - acd\right)}{aC_{22}^S} \right) \overset{0}{M_{22}^{bT}}$$

$$\left. - C_{22}^0 d \left(L^2 + x_2 L - x_2^2\right) \left( \frac{ac}{C_{22}^S} \overset{0}{(M_{22}^{zT} + M_{22}^{ST})} \right) \right\}$$

$$\overset{1}{\varphi_2}(x_2) = \frac{1}{C_{22}^0 d} \frac{L-2x_2}{24}$$ (104)

$$\left\{ \left( \frac{ac(C_{22}^{S0} C_{22}^2 - C_{22}^{1S} C_{22}^1)}{C_{22}^S} \overset{0}{(L^2 + 2x_2 L - 2x_2^2)} + \frac{12\left(a(C_{22}^{S0} C_{22}^2 - C_{22}^{1S} C_{22}^1) + K^2 C_{55}^P C_{22}^1\right)}{aC_{22}^S} \right) N_{22}^T \right.$$

$$+ \left( \frac{ac^2 d C_{22}^0}{C_{22}^S} \overset{0}{(L^2 + 2x_2 L - 2x_2^2)} + \frac{12 C_{22}^0 \left(2acd - K^2 C_{55}^P\right)}{aC_{22}^S} \right) M_{22}^{bT}$$

$$\left. - C_{22}^0 d \left( \left( \frac{ac\overset{0}{(L^2 + 2x_2 L - 2x_2^2)} + 12}{C_{22}^S} \right) (M_{22}^{zT} + M_{22}^{ST}) \right) \right\}$$

**APPENDIX D – 2D thermo-elasticity solutions for multi-layered wide plates with thermally and mechanically imperfect interfaces subjected to steady-state thermal gradients**

Exact solutions have been obtained in [2,23,4] for temperature and stress/strain fields in simply supported, fully bonded, multilayered plates subjected to steady-state thermal loading with sinusoidal in-plane distribution. The derivation is based on the preliminary definition of the three-dimensional temperature field, through the solution of the heat conduction equation in each layer and the imposition of thermal boundary and continuity conditions at the interfaces; this is followed by the solution of Navier's thermo-elasticity equations in each layer and the imposition of interfacial boundary conditions to define the constants of integration.

Here the formulation proposed in [2] is particularized to plates deforming in cylindrical bending, with $x_2 - x_3$ the bending plane, and extended to account for interfaces which may be mechanically and/or thermally imperfect or fully debonded. The plates, of length $L$ and thickness $h$ in the bending plane, are made of orthotropic layers with the material axes parallel to the reference axes (cross-ply layup); the layers have arbitrary thickness, $^{(k)}h$, elastic constants and stacking sequence (Fig. 1). The plane-strain thermo-elastic constitutive equations of the layer $k$ are derived from the 3D thermoelasticity relationships [26]:

$$^{(k)}\begin{Bmatrix} \sigma_{22} \\ \sigma_{33} \\ \sigma_{23} \end{Bmatrix} = {}^{(k)}\begin{bmatrix} C_{22} & C_{23} & 0 \\ C_{23} & C_{33} & 0 \\ 0 & 0 & C_{55} \end{bmatrix} {}^{(k)}\begin{Bmatrix} \varepsilon_{22} - \alpha_2 T \\ \varepsilon_{33} - \alpha_3 T \\ 2\varepsilon_{23} \end{Bmatrix} - {}^{(k)}\begin{Bmatrix} C_{21}\alpha_1 T \\ C_{31}\alpha_1 T \\ 0 \end{Bmatrix}$$ (105)

with $^{(k)}C_{ij}$ (for $i,j = 2,3$) the coefficients of the 3D stiffness matrix in engineering notation, $^{(k)}\alpha_i$ the coefficient of thermal expansion along the $x_i$ principal material direction and $^{(k)}T = {}^{(k)}T(x_2, x_3)$ the temperature increment in the layer $k$. Eq. (105) assumes that the coupling between elastic deformations in the layer and heat transfer can be neglected and that the temperature distribution is prescribed.

The mechanical behavior of the interfaces is described by the interfacial tractions laws, $\hat{\sigma}_S^k(x_2) = [{}^{(k)}\sigma_{23}(x_2, x_3 = x_3^k)]n_3^k\big|_{{}^{(k)}\mathscr{S}^+} = K_S^k \hat{v}_2^k(x_2)$ and $\hat{\sigma}_N^k(x_2) = [{}^{(k)}\sigma_{33}(x_2, x_3 = x_3^k)]n_3^k\big|_{{}^{(k)}\mathscr{S}^+} = K_N^k \hat{v}_3^k(x_2)$ with $n_3^k$ the component of the outward normal to the surface; the laws relates the interfacial tractions to the relative sliding and opening displacements, $\hat{v}_2^k(x_2) = {}^{(k+1)}v_2(x_2, x_3 = x_3^k) - {}^{(k)}v_2(x_2, x_3 = x_3^k)$ and $\hat{v}_3^k(x_2) = {}^{(k+1)}v_3(x_2, x_3 = x_3^k) - {}^{(k)}v_3(x_2, x_3 = x_3^k)$.

Following [3], the thermal behavior of the interface $^{(k)}\mathscr{S}^+$ at the coordinate $x_3 = x_3^k$ is controlled by an interfacial thermal resistance $R^k = 1/H^k$, which is defined as the reciprocal of the interlayer thermal conductance $H^k$ and relate the heat flux through the interface, $^{(k)}q_3^k = -{}^{(k)}K_3 {}^{(k)}T^k{}_{,3}$, to the interfacial temperature variation, $^{(k)}q_3^k = -\frac{1}{R^k}\left[{}^{(k+1)}T^k - {}^{(k)}T^k\right]$, where $^{(k)}q_3^k = {}^{(k)}q_3(x_3 = x_3^k)$ and $^{(k)}T^k = {}^{(k)}T(x_3 = x_3^k)$; when the layers are in perfect thermal contact, $R^k = 0$ and the temperature becomes continuous at the interface, $^{(k)}T^k = {}^{(k+1)}T^k$; when $H^k = 0$, the interface becomes impermeable and there is no heat flux, $^{(k)}q_3^k = 0$.

Exact solutions are found for a simply supported plate with boundary conditions given by $v_3, \sigma_{22} = 0$ at $x_2 = 0, L$. The plate is subjected to applied temperatures onto the upper and lower surfaces $\mathscr{S}^+$ and $\mathscr{S}^-$ at $x_3 = x_3^n$ and $x_3 = x_3^0$:

$$T(x_2, x_3 = x_3^n) = T_{m1} \sin(px_2), \qquad T(x_2, x_3 = x_3^0) = T_{m2} \sin(px_2) \qquad (106)$$

with $p = m\pi/L$ and $m \in \mathbb{N}$ (Fig. 3c). Solutions for applied temperatures other than Eq. (106) can be obtained from the results obtained here using superposition and Fourier's series expansions. A uniform temperature field applied to the upper surface, $T(x_3 = x_3^n) = T_1$, for instance, is approximated using the distribution (106) and the expansion:

$$T(x_2, x_3 = x_3^n) = T_1 \sum_{k=1}^{k\,max} \frac{4}{\pi(2k-1)} \sin\left[(2k-1)\frac{\pi}{L} x_2\right] \quad \text{for } 0 \leq x_2 \leq L \tag{107}$$

For large values of $k_{max}$, the series (107) defines an applied temperature which is approximately constant over the domain but for some boundary regions near $x_2 = 0, L$, whose sizes are proportional to the wave length of $k_{max}$ and decrease on increasing $k_{max}$; there the Fourier sum overshoots by an amount which does not vanishes on increasing $k_{max}$ and for the square half wave of height $T_1$ of Eq. (107) the error in the approximation is around 18%.

**D.1 Heat conduction problem**

The thermal boundary conditions and continuity conditions at the $n$-1 interfaces are:

$$\begin{aligned}
&^{(n)}T(x_2, x_3 = x_3^n) = T_{m1} \sin(px_2) \\
&^{(1)}T(x_2, x_3 = x_3^0) = T_{m2} \sin(px_2) \\
&^{(k)}q_3^k(x_2) = -\frac{1}{R^k}\left[^{(k+1)}T^k(x_2) - {}^{(k)}T^k(x_2)\right], \text{ for } k = 1..n-1 \\
&^{(k)}q_3^k(x_2) = {}^{(k+1)}q_3^k(x_2), \text{ for } k = 1..n-1
\end{aligned} \tag{108}$$

In the special cases of perfect thermal contact and impermeable interfaces, the continuity conditions in Eqs. (108) become:

$$\begin{aligned}
&^{(k)}T^k(x_2) = {}^{(k+1)}T^k(x_2), \quad \text{with } R^k = 0, \text{ for } k = 1..n-1 \\
&^{(k)}q_3^k(x_2) = 0, \quad \text{with } H^k = 0, \text{ for } k = 1..n-1
\end{aligned} \tag{109}$$

The steady state heat conduction equation for a homogeneous orthotropic solid, where all the fields variable depend on $x_2$ and $x_3$ only, is given by:

$$K_2 \frac{\partial^2 T(x_2, x_3)}{\partial x_2^2} + K_3 \frac{\partial^2 T(x_2, x_3)}{\partial x_3^2} = 0 \tag{110}$$

where $K_i$ is the thermal conductivity in the $x_i$ principal material direction which coincides with a geometrical axis. The temperature field:

$$^{(k)}T(x_2, x_3) = {}^{(k)}F(x_3) \sin(px_2) \tag{111}$$

with

$$^{(k)}F(x_3) = {}^{(k)}\left(\overline{c}_1 e^{s_1 x_3} + \overline{c}_2 e^{s_2 x_3}\right)$$

$$^{(k)}s_{1,2} = \pm \sqrt{\frac{{}^{(k)}K_2 p^2}{{}^{(k)}K_3}} \quad \text{and} \quad {}^{(k)}s_2 = -{}^{(k)}s_1 \tag{112}$$

solves the Eq. (110) and the imposition of the $2n$ boundary and continuity conditions, Eqs. (108), leads to an algebraic system of $2n$ equations in the $2n$ unknown constants ${}^{(k)}\overline{c}_{1,2}$, for $k = 1,...,n$ [2]. Since the ${}^{(k)}s_{1,2}$ are real, Eq. (112) can be written as:

$$^{(k)}F(x_3) = {}^{(k)}c_1 \cosh\left({}^{(k)}s_1 x_3\right) + {}^{(k)}c_2 \sinh\left({}^{(k)}s_2 x_3\right) \tag{113}$$

with the $2n$ unknown constants ${}^{(k)}c_1 = {}^{(k)}\overline{c}_1 + {}^{(k)}\overline{c}_2$ and ${}^{(k)}c_2 = {}^{(k)}\overline{c}_1 - {}^{(k)}\overline{c}_2$ for $k = 1,...,n$.

If the applied temperatures are uniform and described by the Fourier expansion, Eq. (107), the temperature field in each layer is approximated over most of the plate domain, away from the boundaries at $x_2 = 0, L$, by

$$^{(k)}T(x_2, x_3) = \sum_{j=1}^{jmax} \frac{4}{m\pi} \left[ {}^{(k)}c_1 \cosh\left({}^{(k)}s_1 x_3\right) + {}^{(k)}c_2 \sinh\left({}^{(k)}s_2 x_3\right) \right] \sin\left(\frac{m\pi}{L} x_2\right) \tag{114}$$

$$m = 2j - 1$$

The temperature distribution in the plate will depend on the relative thermal conductivity of the layers, the interface thermal resistance and the assigned temperature distribution. The sinusoidal temperature of Eq. (106) originates piecewise nonlinear temperature fields, which become piecewise linear in the central portion of the plate when $p = m\pi/L$ is small, e.g. in thin plates. If the applied temperatures are uniform along $x_2$, $T(x_2, x_3 = x_3^n) = T_1$ and $T(x_2, x_3 = x_3^0) = T_2$, the temperature distribution in the layers is piece-wise linear, Eq. (110), with jumps at the interfaces which are not in perfect thermal contact; the slope of the pieces varies between layers characterized by different values of thermal conductivity $^{(k)}K_3$ or by interfacial thermal contact resistance. In a classical cross-ply laminate made of $n$ unidirectionally reinforced plies of equal thickness, $^{(k)}h = h/n$, where $^{(k)}K_3 = K_3$ is the same in all layers, and thermal contact is imperfect, the temperature distribution in the layer $k$ is:

$$^{(k)}T(x_3) = {}^{(k)}c_1 + {}^{(k)}c_2 \frac{(x_3 - x_3^0)}{h} \tag{115}$$

with

$$^{(k)}c_1 = T_2 + \frac{T_1 - T_2}{1 + K_3 \sum_{i=1}^{n-1} \frac{R^i}{h}} K_3 \sum_{i=1}^{k-1} \frac{R^i}{h}, \quad {}^{(k)}c_2 = \frac{T_1 - T_2}{1 + K_3 \sum_{i=1}^{n-1} \frac{R^i}{h}}. \tag{116}$$

### D.2 Thermo-elastic problem

Using the thermo-elastic constitutive equations (105) and the compatibility equations relating strain and displacements, the equilibrium equations for the generic layer $k$ are given in terms of the displacement variables, $^{(k)}v_2$ and $^{(k)}v_3$, by Navier's equations:

$$^{(k)}C_{22}{}^{(k)}v_{2,22} + {}^{(k)}(C_{23} + C_{55}){}^{(k)}v_{3,23} + {}^{(k)}C_{55}{}^{(k)}v_{2,33} = {}^{(k)}(C_{12}\alpha_1 + C_{22}\alpha_2 + C_{23}\alpha_3){}^{(k)}T(x_2, x_3)_{,2}$$
$$^{(k)}C_{33}{}^{(k)}v_{3,33} + {}^{(k)}(C_{23} + C_{55}){}^{(k)}v_{2,23} + {}^{(k)}C_{55}{}^{(k)}v_{3,22} = {}^{(k)}(C_{13}\alpha_1 + C_{23}\alpha_2 + C_{33}\alpha_3){}^{(k)}T(x_2, x_3)_{,3} \tag{117}$$

The $2n$ differential equations, in the $2n$ variables, are complemented by $4n$ boundary and continuity conditions. The boundary conditions are:

$$^{(n)}\sigma_{23}(x_2, x_3^n) = {}^{(n)}\sigma_{33}(x_2, x_3^n) = {}^{(1)}\sigma_{23}(x_2, x_3^0) = {}^{(1)}\sigma_{33}(x_2, x_3^0) = 0 \tag{118}$$

and the continuity conditions at the imperfect interfaces, which impose traction continuity and relative jumps controlled by the interfacial traction laws, are:

$${}^{(k)}\sigma_{23}(x_2, x_3^k) = {}^{(k+1)}\sigma_{23}(x_2, x_3^k)$$

$${}^{(k)}\sigma_{33}(x_2, x_3^k) = {}^{(k+1)}\sigma_{33}(x_2, x_3^k)$$

$${}^{(k)}\sigma_{23}(x_2, x_3^k) = K_S^k \left[ {}^{(k+1)}v_2(x_2, x_3^k) - {}^{(k)}v_2(x_2, x_3^k) \right],$$

$${}^{(k)}\sigma_{33}(x_2, x_3^k) = K_N^k \left[ {}^{(k+1)}v_3(x_2, x_3^k) - {}^{(k)}v_3(x_2, x_3^k) \right]$$

for $k = 1..n-1$ (119)

The displacement variables in the *k*th layer are defined by the sum of a complementary and a particular solution, ${}^{(k)}v_2(x_2, x_3) = {}^{(k)}v_{2c}(x_2, x_3) + {}^{(k)}v_{2p}(x_2, x_3)$ and ${}^{(k)}v_3(x_2, x_3) = {}^{(k)}v_{3c}(x_2, x_3) + {}^{(k)}v_{3p}(x_2, x_3)$.

### D.2.1 Particular solution

A particular solution of the system (117), with ${}^{(k)}T(x_2, x_3)$ given by Eqs. (111) and (113), which satisfies the boundary conditions, is:

$${}^{(k)}v_{2p}(x_2, x_3) = {}^{(k)}\left( B_1 e^{s_1 x_3} + B_2 e^{-s_1 x_3} \right) \cos(px_2)$$

$${}^{(k)}v_{3p}(x_2, x_3) = {}^{(k)}\left( D_1 e^{s_1 x_3} + D_2 e^{-s_1 x_3} \right) \sin(px_2)$$ (120)

$$p = m\pi / L \text{ and } {}^{(k)}s_1 = +\sqrt{\frac{{}^{(k)}K_2 p^2}{{}^{(k)}K_3}}$$

where ${}^{(k)}B_1, {}^{(k)}B_2, {}^{(k)}D_1, {}^{(k)}D_2$ are unknown constants. Substituting Eqs. (120) into Eqs. (117) and collecting the terms multiplying $e^{s_1 x_3}$ and $e^{-s_1 x_3}$, leads to four algebraic equations in the unknown ${}^{(k)}B_1, {}^{(k)}B_2, {}^{(k)}D_1, {}^{(k)}D_2$:

$${}^{(k)}\left[ C_{22}p^2 B_1 - (C_{23} + C_{55}) ps_1 D_1 - C_{55}s_1^2 B_1 + pc_1 (C_{12}\alpha_1 + C_{22}\alpha_2 + C_{23}\alpha_3) \right] = 0$$

$${}^{(k)}\left[ C_{22}p^2 B_2 + (C_{23} + C_{55}) ps_1 D_2 - C_{55}s_1^2 B_2 + pc_2 (C_{12}\alpha_1 + C_{22}\alpha_2 + C_{23}\alpha_3) \right] = 0$$

$${}^{(k)}\left[ C_{33}s_1^2 D_1 - (C_{23} + C_{55}) ps_1 B_1 - C_{55}p^2 D_1 - s_1 c_1 (C_{13}\alpha_1 + C_{23}\alpha_2 + C_{33}\alpha_3) \right] = 0$$ (121)

$${}^{(k)}\left[ C_{33}s_1^2 D_2 + (C_{23} + C_{55}) ps_1 B_2 - C_{55}p^2 D_2 + s_1 c_2 (C_{13}\alpha_1 + C_{23}\alpha_2 + C_{33}\alpha_3) \right] = 0$$

**D.2.2 Complementary solution**

An appropriate solution of the complementary problem, which satisfies the simply support conditions is:

$$^{(k)}v_{2c}(x_2, x_3) = {}^{(k)}V(x_3)\cos(px_2)$$
$$^{(k)}v_{3c}(x_2, x_3) = {}^{(k)}W(x_3)\sin(px_2) \tag{122}$$
$$p = m\pi/L$$

with

$$\left[{}^{(k)}V(x_3), {}^{(k)}W(x_3)\right] = \left[{}^{(k)}V_0, {}^{(k)}W_0\right]e^{tx_3} \tag{123}$$

Substituting Eqs. (123) into the homogeneous part of Eqs. (117) yields a homogeneous system of algebraic equations whose non-trivial solution is defined by the characteristic equation [24]:

$$A_0 t^4 + A_1 t^2 + A_2 = 0 \tag{124}$$

with

$$A_0 = C_{33}C_{55}, \quad A_1 = p^2\left[(C_{23}+C_{55})^2 - C_{22}C_{33} - C_{55}^2\right], \quad A_2 = p^4 C_{22}C_{55} \tag{125}$$

Eq. (124) can be written in the form of a quadratic equation with

$$A_0\gamma^2 + A_1\gamma + A_2 = 0$$
$$\gamma = t^2 \tag{126}$$
$$\Delta = A_1^2 - 4A_0 A_2$$

The nature of the solution depends on the elastic constants of the layer through the discriminant $\Delta$. In [2] only solutions corresponding to positive discriminants were examined, since they correspond to typical orthotropic layers used in laminated systems. Here, following what done in [24,30], solutions are presented also for the case of zero discriminant, which describe isotropic layers, and for negative discriminants [24], which could describe orthotropic layers which are stiffer in the transverse direction, e.g. sandwich cores.

*Positive discriminant*

When $\Delta$ is positive, Eq. (126) has two real and unequal roots, $\gamma_{1,2} = (-A_1 \pm \sqrt{\Delta})/2A_0$. By setting $m_j = \sqrt{|\gamma_j|}$ for $j=1,2$, if $\gamma_j > 0$ then the roots of Eq. (126) are real, $t_j = \pm m_j$, so that:

$$^{(k)}V(x_3) = \sum_{j=1}^{2} a_{1j} \cosh(m_j x_3) + a_{2j} \sinh(m_j x_3) \qquad (127)$$

$$^{(k)}W(x_3) = \sum_{j=1}^{2} b_{1j} \cosh(m_j x_3) + b_{2j} \sinh(m_j x_3)$$

where four of the eight constants are derived substituting (127) into the equilibrium equations (117),

$$b_{1j} = \beta_j a_{2j}, \quad b_{2j} = \beta_j a_{1j},$$

$$\beta_j = \frac{(C_{23} + C_{55}) p t_j}{C_{33} t_j^2 - C_{55} p^2} \qquad (128)$$

(formulas (128) corrects a sign omission in the derivation in [24]).

If $\gamma_j < 0$, the roots of Eq. (126) are complex and $t_j = \pm i m_j$ so that:

$$^{(k)}V(x_3) = \sum_{j=1}^{2} a_{1j} \cos(m_j x_3) + a_{2j} \sin(m_j x_3) \qquad (129)$$

$$^{(k)}W(x_3) = \sum_{j=1}^{2} b_{1j} \cos(m_j x_3) + b_{2j} \sin(m_j x_3)$$

Where four of the eight constants are derived substituting (129) into the equilibrium equations (117),

$$b_{1j} = -\alpha_j a_{2j}, \quad b_{2j} = \alpha_j a_{1j},$$

$$\alpha_j = \frac{(C_{23} + C_{55}) p t_j}{C_{33} t_j^2 + C_{55} p^2} \qquad (130)$$

The remaining four constants for each layer are determined through boundary and continuity conditions (119).

*Zero discriminant*

When the layer is isotropic, $\Delta = 0$, Eq. (126) has two real and equal roots, $\gamma_{1,2} = -A_1/2A_0 = p^2$ and

Eq. (124) has two repeated roots $t_j = \pm p$, for $j=1,2$. The displacement functions are:

$$^{(k)}V(x_3) = (a_1 + c_1 x_3) e^{px_3} + (a_2 + c_2 x_3) e^{-px_3}$$
$$^{(k)}W(x_3) = (b_1 + d_1 x_3) e^{px_3} + (b_2 + d_2 x_3) e^{-px_3} \tag{131}$$

where four of the eight constants are derived substituting (131) into the equilibrium equations (117):

$$d_1 = c_1, \quad d_2 = -c_2,$$
$$b_1 = a_1 - \left[\frac{2C_{55}}{(C_{23}+C_{55})p} + \frac{1}{p}\right] c_1, \quad b_{2j} = -a_2 - \left[\frac{2C_{55}}{(C_{23}+C_{55})p} + \frac{1}{p}\right] c_2, \tag{132}$$

The remaining four constants for each layer are determined through boundary and continuity conditions (119).

*Negative discriminant*

When $\Delta < 0$, Eq. (126) has two complex conjugate roots, $\gamma_{1,2} = \mu_r \pm i\mu_c = r(\cos\theta \pm i\sin\theta)$, where $\mu_r = -A_1/2A_0$ and $\mu_c = \sqrt{|\Delta|}/2A_0$, $r = \sqrt{\mu_c^2 + \mu_r^2}$ and $\theta = \arctan(\mu_c/\mu_r)$. The corresponding roots of Eq. (126) are $t_{1,2,3,4} = \pm(\rho_1 \pm i\rho_2)$, where $\rho_1 = \sqrt{r}\cos(\theta/2)$ and $\rho_2 = \sqrt{r}\sin(\theta/2)$. The displacement functions are:

$$^{(k)}V(x_3) = a_1 e^{\rho_1 x_3} \cos(\rho_2 x_3) + a_2 e^{\rho_1 x_3} \sin(\rho_2 x_3) + a_3 e^{-\rho_1 x_3} \cos(\rho_2 x_3) + a_4 e^{-\rho_1 x_3} \sin(\rho_2 x_3)$$
$$^{(k)}W(x_3) = b_1 e^{\rho_1 x_3} \cos(\rho_2 x_3) + b_2 e^{\rho_1 x_3} \sin(\rho_2 x_3) + b_3 e^{-\rho_1 x_3} \cos(\rho_2 x_3) + b_4 e^{-\rho_1 x_3} \sin(\rho_2 x_3) \tag{133}$$

where four of the eight constants are derived substituting Eq. (133) into the equilibrium equations (117):

$$b_1 = r_1 a_1 - r_2 a_2, \quad b_2 = r_2 a_1 + r_1 a_2,$$
$$b_3 = -r_1 a_3 - r_2 a_4, \quad b_4 = r_2 a_3 - r_1 a_4 \tag{134}$$

where

$$r_1 = \frac{\rho_1 \left[C_{22} p^2 - C_{55}(\rho_1^2 + \rho_2^2)\right]}{p(C_{23}+C_{55})(\rho_1^2 + \rho_2^2)}, \quad r_2 = \frac{\rho_2 \left[C_{22} p^2 + C_{55}(\rho_1^2 + \rho_2^2)\right]}{p(C_{23}+C_{55})(\rho_1^2 + \rho_2^2)} \tag{135}$$

The remaining four constants are determined through boundary and continuity conditions (119).